\documentclass[%
preprint,
 amsmath,amssymb,
 aps,
prstab,
]{revtex4-2}

\usepackage{graphicx}
\usepackage{dcolumn}
\usepackage{bm}

\begin{document}


\title{Wake fields and impedance of space charge in Cartesian coordinate system}

\author{Demin Zhou}
 \altaffiliation[Also at ]{Graduate University for Advanced Studies (SOKENDAI)}
 \email{dmzhou@post.kek.jp}
\affiliation{%
KEK, High Energy Accelerator Organization, Oho 1-1, Tsukuba 305-0801, Japan
}%
\author{Yuancun Nie}
 \email{nieyuancun@whu.edu.cn}
 \affiliation{The Institute for Advanced Studies, Wuhan University, Wuhan 430072, Hubei, P.R. China}

\date{\today}

\begin{abstract}
In Cartesian coordinate system, the fields generated by a point charge moving parallel to the axis of a rectangular vacuum chamber can be formulated in terms of eigenfunctions of the rectangular waveguide using the mode expansion method. In combination with the conventional impedance theory, the Green-function forms of the wake functions and impedance for space-charge effects can be obtained, and are found to be functions of the positions of both the source and test particles. Using the Green's functions calculated for a point charge, the wake fields and impedance of a beam with various distributions can be calculated, and should be useful to model the three-dimensional space-charge effects. This paper summarizes our findings and also show comparisons to the existing theories.
\end{abstract}

\maketitle


\section{\label{sec:introduction}Introduction}

Space charge is an important source of collective effects and can exert a strong impact on the machine performance of modern particle accelerators with low-energy but high-intensity beams, or with high-energy but high peak-current beams. To simulate the beam dynamics with space charge, many tracking codes use self-consistent models based on particle-in-cell (PIC) method (for typical examples, see Refs.~\cite{Qiang2000434,GPTwebsite}). On the other hand, many other codes use non-self-consistent models based on space charge impedance with given beam distribution (for an example, see Ref.~\cite{BorlandelegantCode}). 

The analytic theories of space charge wake functions and impedance have usually been derived in the cylindrical coordinate system with round vacuum chambers~\cite{Alex1993,NgBook2006,Shobuda:2018bqc}. With the wake potential expanded in terms of cylindrical coordinates, usually the leading terms of longitudinal monopole and the transverse dipole are of important concern for evaluations of space charge effects~\cite{Ng:2004ky}. In recent years, efforts have been made to extend the theories to cover various cases such as non-round chambers or non-uniform beam distributions. In Ref.~\cite{WangPRSTAB2015}, the longitudinal space charge impedance of a round uniform beam was studied in details in the presence of parallel-plates or rectangular chambers. In Refs.~\cite{PersichelliPRAB2017,MiglioratiPRAB2018}, the authors looked into the case of chambers with elliptic geometry for which the Mathieu functions can be used to express the electromagnetic fields.

In this paper, the theory of space charge wake fields and impedance is discussed in the Cartesian coordinate system. To preserve the generality of the theory, the traditional treatment of expanding the wake potential in transverse directions is not used. Therefore, the wake functions and impedance formulated in this paper are functions of transverse coordinates. The reader will note that such formulations provide the convenience of modeling space charge effects for three-dimensional beam distributions as well as three-dimensional chamber geometries.

The content of the paper is organized as follows. In Sec.~\ref{sec:Fundamental_Formulations} we formulate the fundamental equations required for the approach of wake fields and impedance in Cartesian coordinate system. In Sec.~\ref{sec:Mode_Expansion_Method} we briefly introduce the mode expansion method applied to obtain the Green's function solution for the vector potential in terms of eigenmodes of rectangular chambers. The explicit forms of space charge wake functions and impedance of a point charge in the rectangular chamber and in free space are derived in Sec.~\ref{sec:Point_Charge_Impedance}. The longitudinal and transverse wake fields and impedance are derived in Sec.~\ref{sec:Longitudinal-wake-with-bunch-distribution} and Sec.~\ref{sec:Transverse-wake-with-bunch-distribution}, respectively. Finally, we summarize our findings and give concluding remarks in Sec.~\ref{sec:Summary}.

\section{\label{sec:Fundamental_Formulations}Fundamental formulations for wake fields and impedance}

In particle accelerators the charged beam generates electromagnetic fields when traveling inside the vacuum chamber. The beam-induced fields are usually referred to as wake fields in the literature. A general formulation of wake fields and their corresponding Fourier transforms (i.e. impedance) in Cartesian coordinate system is given in Ref.~\cite{Palumbo1994}, and it forms the basis of discussions in this paper.

Consider a charged particle moving in parallel to the axis of the vacuum chamber  with constant velocity $\vec{v}=\vec{i}_zv$. With the charge density defined by Dirac delta functions as
\begin{equation}
\varrho(\vec{R},t)
=
q_0
\delta(x-x_0)
\delta(y-y_0)
\delta(z-z_0)
\label{eq:PointChargeDensity1}
\end{equation}
where $z_0\equiv vt$ and $\vec{R}\equiv (x,y,z)$ the spatial vector position, the current density is given by $\vec{\mathcal{J}}(\vec{R},t)=\varrho(\vec{R},t)\vec{v}$. One can apply them to the Maxwell's equations with boundary conditions and obtain the time-varying electromagnetic fields $\vec{\mathcal{E}}(\vec{R},t)$ and $\vec{\mathcal{B}}(\vec{R},t)$.

Suppose that a test charged particle $q_1$ with coordinates $\vec{R}_1=(x_1,y_1,z_1)$ follows $q_0$ with the same velocity $v$ but at a time delay of $\tau =z/v=(z_0-z_1)/v$. The Lorentz force acted on $q_1$ is then given by
\begin{equation}
\vec{\mathcal{F}}(\vec{R}_1,\vec{R}_0;t)
=
q_1
\left[
\vec{\mathcal{E}}(\vec{R}_1,\vec{R}_0;t)
+
\vec{v}
\times
\vec{\mathcal{B}}(\vec{R}_1,\vec{R}_0;t)
\right]
.
\label{eq:LorentzForce1}
\end{equation}
With the rigid-beam and impulse approximations~\cite{ChaoNotes2002}, the impulse kick exerted on $q_1$ when it travels by a length of $L$ is calculated by integrating the Lorentz force as
\begin{equation}
\overline{\vec{\mathcal{F}}}
(\vec{r}_{1},\vec{r}_{0};\tau)
=
\int_{\tau}^{\tau+\frac{L}{v}}
dt\
v
\vec{\mathcal{F}}
(\vec{R}_1,\vec{R}_0;t)
.
\quad
\label{eq:WakePotentialDefinition1}
\end{equation}
The vectors $\vec{r}_{1}$ and  $\vec{r}_{0}$ in Eq.~(\ref{eq:WakePotentialDefinition1}) represent the transverse positions of the test and source particles respectively, i.e. $\vec{r}_{1}=(x_1,y_1)$ and $\vec{r}_{0}=(x_0,y_0)$. The quantity $\overline{\vec{\mathcal{F}}}=(\overline{\mathcal{F}}_x,\overline{\mathcal{F}}_y,\overline{\mathcal{F}}_z)$ is called the wake potential, which is a function of $\tau$ and the transverse coordinates of source and test particles. Then the wake functions in the three dimensions are defined as follows
\begin{subequations}
\begin{equation}
w_z(\vec{r}_{1},\vec{r}_{0};\tau)
=
-
\frac{1}{q_0q_1}
\overline{\mathcal{F}}_z
(\vec{r}_{1},\vec{r}_{0};\tau)
,
\label{eq:LongitudinalWakeFunctionDefinition1}
\end{equation}
\begin{equation}
w_\perp(\vec{r}_{1},\vec{r}_{0};\tau)
=
\frac{1}{q_0q_1}
\overline{\mathcal{F}}_\perp
(\vec{r}_{1},\vec{r}_{0};\tau)
,
\label{eq:TransverseWakeFunctionDefinition1}
\end{equation}
\label{eq:WakeFunctionDefinition1}
\end{subequations}
with $\perp$ indicating $x$ or $y$. Here we use a minus sign for the longitudinal wake function by following the conventional definition of longitudinal wake functions. Using Fourier transform, one can calculate the spectrum of the wake functions, so called impedance, as
\begin{subequations}
\begin{equation}
Z_\parallel(\vec{r}_{1},\vec{r}_{0};\omega)
=
\int_{-\infty}^{\infty}
d\tau \
w_z (\vec{r}_{1},\vec{r}_{0};\tau)
e^{i\omega \tau}
,
\label{eq:LongitudinalImpedanceDefinition1}
\end{equation}
\begin{equation}
Z_\perp(\vec{r}_{1},\vec{r}_{0};\omega)
=
\kappa
\int_{-\infty}^{\infty}
d\tau \
w_\perp (\vec{r}_{1},\vec{r}_{0};\tau)
e^{i\omega \tau}
,
\label{eq:TransverseImpedanceDefinition1}
\end{equation}
\label{eq:ImpedanceDefinition1}
\end{subequations}
with the constant $\kappa$ defined in a conventional way as $\kappa=i/\beta$ and $\beta=v/c$ the relative velocity~\cite{NgBook2006}. Then the wake functions expressed by inverting the above Fourier transforms are
\begin{subequations}
\begin{equation}
w_z(\vec{r}_{1},\vec{r}_{0};\tau)
=
\frac{1}{2\pi}
\int_{-\infty}^{\infty}
d\omega \
Z_\parallel (\vec{r}_{1},\vec{r}_{0};\omega)
e^{-i\omega \tau}
,
\label{eq:LongitudinalWakeFunctionByImpedance1}
\end{equation}
\begin{equation}
w_\perp(\vec{r}_{1},\vec{r}_{0};\tau)
=
\frac{1}{2\pi\kappa}
\int_{-\infty}^{\infty}
d\omega \
Z_\perp (\vec{r}_{1},\vec{r}_{0};\omega)
e^{-i\omega \tau}
.
\label{eq:TransverseWakeFunctionByImpedance1}
\end{equation}
\label{eq:WakeFunctionByImpedance1}
\end{subequations}
The reader may notice that the constant $\kappa$ in Eqs.~(\ref{eq:TransverseImpedanceDefinition1}) and~(\ref{eq:TransverseWakeFunctionByImpedance1}) can be any value. For example, $\kappa=i$ is also frequently used in the literature.

The above formulation of wake fields and impedance is very general and is applicable to the cases of vacuum chambers with arbitrary shapes. If the chamber considered is cylindrically symmetric, the whole theory can be discussed under the framework of cylindrical coordinate system. This is relevant to the classical theory as shown in Ref.~\cite{Alex1993}, but it is beyond the scope of this paper.

The wake functions and impedance driven by a point charge as formulated in this section can be used as Green's functions to calculate the wake potentials and coupling impedance with beam distributions. Let us define $w$ and $W$ as the wake functions of a point charge and a bunch distribution, respectively. Correspondingly, $Z$ and $\mathcal{Z}$ are the impedance of a point charge and a bunch distribution, respectively. The impedance of a bunch is calculated by integration with respect to the distribution function as
\begin{equation}
	\mathcal{Z}_u(\vec{r};k)
	=
	\int_{-\infty}^{\infty}dx'\int_{-\infty}^{\infty}dy'
	\rho_\perp(x',y')
	Z_u(\vec{r},\vec{r}';k)
	\label{eq:Impedance_Bunch_General1}
\end{equation}
where $u$ represents $x$, $y$, or $\parallel$, and wake functions of a bunch as
\begin{equation}
	W_u(\vec{r};z)
	=
	\int_{-\infty}^{\infty}dx'\int_{-\infty}^{\infty}dy' \int_{-\infty}^{\infty}dz'
	\rho(x',y',z')
	w_u(\vec{r},\vec{r}';z-z')
	\label{eq:Wake_Potential_Bunch_General1}
\end{equation}
where $u$ represents $x$, $y$, or $z$. Here it is assumed that the bunch distribution is normalized to unity with $\int \rho_\perp(x',y')dx'dy'=1$ and $\int \rho(x',y',z')dx'dy'dz'=1$. Sometimes the averages of impedance and wake potentials over the transverse density are also useful as an approximation by reducing the model from three-dimensional to one-dimensional:
\begin{equation}
	\overline{\mathcal{Z}}_u(k)
	=
	\int_{-\infty}^{\infty}dx\int_{-\infty}^{\infty}dy
	\rho_\perp(x,y)
	\mathcal{Z}_u(\vec{r};k)
	,
	\label{eq:Impedance_Average_Bunch_General1}
\end{equation}
\begin{equation}
	\overline{W}_u(z)
	=
	\int_{-\infty}^{\infty}dx\int_{-\infty}^{\infty}dy
	\rho_\perp(x,y)
	W_u(\vec{r};z)
	.
	\label{eq:Wake_Potential_Average_Bunch_General1}
\end{equation}

\section{\label{sec:Mode_Expansion_Method}Mode expansion method with rectangular chamber}

In this section, we briefly introduce the mode expansion method (also called Ohm-Rayleigh method in the literature) as applied to solving inhomogeneous partial differential equations governing the space charge fields. For more details the reader can refer to Ref.~\cite{ZhouThesis2011}. We start from considering a rectangular waveguide with its transverse dimensions in the region of $-a/2<x<a/2$ and $-b/2<y<b/2$. The chamber has infinite length in the $z$ direction with its walls assumed to be perfectly conductive. The main task is to find the explicit forms of Eqs.~(\ref{eq:LongitudinalImpedanceDefinition1}) and~(\ref{eq:TransverseImpedanceDefinition1}) in terms of eigenmodes of the rectangular waveguide. For a passive waveguide, the delta function of $z$ in Eq.~(\ref{eq:PointChargeDensity1}) can be replaced by its Fourier transform as
\begin{equation}
\delta(z-vt)
=
\frac{1}{2\pi}
\int_{-\infty}^\infty
e^{-ik(z-vt)}\
dk
.
\label{eq:DiracDeltaFourier1}
\end{equation}
The delta function of transverse coordinates can be expanded into the summation of the eigenmodes of the rectangular waveguide as follows
\begin{equation}
\delta(\vec{r}-\vec{r}_{0})
=
\sum_{m=0}^\infty
\sum_{n=0}^\infty
\phi _{mn\nu}\left(\vec {r}\right)
\phi _{mn\nu}\left(\vec {r}_{0}\right)
,
\label{eq:DiracDeltaFunctionExpansion2}
\end{equation}
where $\nu$ represents $x$, $y$, or $z$. And the complete set of orthonormal eigenfunction for the $x$, $y$ and $z$ directions are 
\begin{subequations}
\begin{equation}
\phi_{mnx}\left(\vec {r}\right)
=
\frac{2}{\sqrt{\left(1+\delta _{m0}\right)a b}}
C_x(x)
S_y(y)
\label{eq:eigen2a},
\end{equation}
\begin{equation}
\phi _{mny}\left(\vec {r}\right)
=
\frac{2}{\sqrt{\left(1+\delta _{n0}\right)a b}}
S_x(x)
C_y(y)
\label{eq:eigen2b},
\end{equation}
\begin{equation}
\phi _{mnz}\left(\vec {r}\right)
=
\frac{2}{\sqrt{a b}}
S_x(x)
S_y(y)
\label{eq:eigen2c}
,
\end{equation}
\end{subequations}
where $\delta_{m0}$ and $\delta_{n0}$ are Kronecker deltas. Here we define $C_x(x) \equiv \cos \left(k_x (x+a/2)\right)$, $S_x(x) \equiv \sin \left(k_x (x+a/2)\right)$, $C_y(y) \equiv \cos \left(k_y (y+b/2)\right)$, and $S_y(y) \equiv \sin \left(k_y (y+b/2)\right)$ with the transverse wave numbers $k_x=m\pi/a$ and $k_y=n\pi/b$.

The problem to be solved first is to find the solutions of two inhomogeneous Helmholtz equations for vector and scalar potentials of
\begin{equation}
\nabla^2 \vec{A}
+
k^2
\vec{A}
=
-\mu_0
\vec{J}
\label{eq:VectorHelmholtzEquationFreq1}
\end{equation}
and
\begin{equation}
\nabla^2 \Phi
+
k^2
\Phi
=
-
\frac{\rho}{\epsilon_0}
,
\label{eq:ScalarHelmholtzEquationFreq1}
\end{equation}
in the frequency domain under the Lorentz gauge condition of
$
\Phi
=
\frac{c^2}{i\omega}
\nabla \cdot
\vec{A}
$
.
Here we define $k\equiv \omega/c$, and the quantities $\vec{J}$ and $\rho$ are respectively the Fourier transforms of $\mathcal{\vec{J}}$ and $\varrho$ as follows
\begin{subequations}
\begin{equation}
\rho(\vec{r},k)
=
\frac{q_0}{\beta c}
\delta(x-x_0)
\delta(y-y_0)
e^{ik z/\beta}
,
\label{eq:PointChargeDensitySpectrum1}
\end{equation}
\begin{equation}
\vec{J}(\vec{r},k)
=
\rho(\vec{r},k)\vec{v}
=
\vec{i}_z
q_0
\delta(x-x_0)
\delta(y-y_0)
e^{ik z/\beta}
,
\label{eq:PointChargeCurrentSpectrum1}
\end{equation}
\label{eq:PointChargeSpectrum1}
\end{subequations}
with $\beta$ the relative velocity.

The magnetic induction and electric field are given by
\begin{subequations}
	\begin{equation}
	\vec{B}
	=
	\nabla \times
	\vec{A}
	,
	\label{eq:MagneticInductionFreqDomain1}
	\end{equation}
	\begin{equation}
	\vec{E}
	=
	ikc \vec{A}
	-
	\nabla \Phi
	=
	ikc \vec{A}
	-
	\frac{c}{ik}
	\nabla \nabla \cdot \vec{A}
	.
	\label{eq:ElectricFieldFreqDomain1}
	\end{equation}
\end{subequations}
With the delta functions in Eqs.~(\ref{eq:PointChargeSpectrum1}) substituted by Eq.~(\ref{eq:DiracDeltaFunctionExpansion2}) and then applied to Eq.~(\ref{eq:VectorHelmholtzEquationFreq1}), the vector potential can be formulated as~\cite{ZhouThesis2011}
\begin{equation}
\vec {A}\left(\vec{r},\vec{r}_0;k\right)
=
\mu _0 q_0
\beta^2\gamma^2
\vec{i}_z
\sum _{m,n\geq 0} 
\frac{
   \phi _{mnz}\left(\vec{r}\right)
   \phi _{mnz}\left(\vec{r}_{0}\right)}
{k^2+\beta^2\gamma^2k_c^2}
e^{i k z/\beta}
,
\label{eq:PointChargeVectorPotential4}
\end{equation}
with Lorentz factor $\gamma=1/\sqrt{1-\beta^2}$, and $k_c=\sqrt{k_x^2+k_y^2}$ is the cut-off wave number of rectangular chamber. Equation~(\ref{eq:PointChargeVectorPotential4}) can be achieved by utilizing the dyadic Green's function of the vector potential~\cite{ZhouThesis2011}, or by direct expansion of the vector potential in terms of the waveguide's eigenfunctions. This formulation of vector potential automatically satisfies the boundary conditions for rectangular chambers, and is valid for arbitrary velocity. M. Hess \mbox{\emph{et al.\ }} presented similar formulation of space-charge fields in Ref.~\cite{HessPRSTAB2007} as Eq.~(\ref{eq:PointChargeVectorPotential4}) but they solved the problem in the time domain.

\section{\label{sec:Point_Charge_Impedance}Wake functions and impedance of a point charge}

Equation~(\ref{eq:PointChargeVectorPotential4}) can be applied to the formulations in Sec.~\ref{sec:Fundamental_Formulations} to calculate the wake fields and then the wake functions and impedance of a point charge. With the detailed derivations given in Ref.~\cite{ZhouThesis2011}, here we only give the explicit form of impedance per unit length:
\begin{subequations}
\begin{equation}
\frac{Z_\parallel(\vec{r}_{1},\vec{r}_{0};k)}{L}
=
\frac{4iZ_0k}{ab}
\sum _{m,n\geq 0} 
\frac{
   \phi' _{mnz}\left(\vec{r}_{1}\right)
   \phi' _{mnz}\left(\vec{r}_{0}\right)}
{k^2+\beta^2\gamma^2k_c^2}
,
\label{eq:PointChargeLongImpedanceInWaveguide1}
\end{equation}
\begin{equation}
\frac{Z_x(\vec{r}_{1},\vec{r}_{0};k)}{L}
=
\frac{-4Z_0\beta\kappa}{ab}
\sum _{m,n\geq 0} 
\frac{
   k_x
   \phi' _{mnx}\left(\vec{r}_{1}\right)
   \phi' _{mnz}\left(\vec{r}_{0}\right)}
{k^2+\beta^2\gamma^2k_c^2}
,
\label{eq:PointChargeXImpedanceInWaveguide1}
\end{equation}
\begin{equation}
\frac{Z_y(\vec{r}_{1},\vec{r}_{0};k)}{L}
=
\frac{-4Z_0\beta \kappa}{ab}
\sum _{m,n\geq 0} 
\frac{
   k_y
   \phi' _{mny}\left(\vec{r}_{1}\right)
   \phi' _{mnz}\left(\vec{r}_{0}\right)}
{k^2+\beta^2\gamma^2k_c^2}
,
\label{eq:PointChargeYImpedanceInWaveguide1}
\end{equation}
\label{eq:PointChargeImpedanceInWaveguide1}
\end{subequations}
where $Z_0=\mu_0 c$ is the impedance of vacuum, and the unnormalized eigenfunctions are defined as
\begin{subequations}
\begin{equation}
\phi '_{mnx}\left(\vec{r}\right)
=
C_x(x)
S_y(y)	
\label{eq:eigen3a},
\end{equation}
\begin{equation}
\phi '_{mny}\left(\vec{r}\right)
=
S_x(x)
C_y(y)
\label{eq:eigen3b},
\end{equation}
\begin{equation}
\phi '_{mnz}\left(\vec{r}\right)
=
S_x(x)
S_y(y)
\label{eq:eigen3c}.
\end{equation}
\label{eq:engen3}
\end{subequations}
The wake functions corresponding to Eqs.~(\ref{eq:PointChargeImpedanceInWaveguide1}) are
\begin{subequations}
\begin{equation}
\frac{w_z(\vec{r}_{1},\vec{r}_{0};z)}{L}
=
\frac{2Z_0c}{ab}
\text{sgn}(z)
\sum _{m,n\geq 0} 
   \phi' _{mnz}\left(\vec{r}_{1}\right)
   \phi' _{mnz}\left(\vec{r}_{0}\right)
e^{-\gamma k_c |z|}
,
\label{eq:PointChargeLongWakeFunctionInWaveguide1}
\end{equation}
\begin{equation}
\frac{w_x(\vec{r}_{1},\vec{r}_{0};z)}{L}
=
\frac{-2Z_0c}{\gamma ab}
\sum _{m,n\geq 0} 
   \frac{k_x}{k_c}
   \phi' _{mnx}\left(\vec{r}_{1}\right)
   \phi' _{mnz}\left(\vec{r}_{0}\right)
e^{-\gamma k_c |z|}
,
\label{eq:PointChargeXWakeFunctionInWaveguide1}
\end{equation}
\begin{equation}
\frac{w_y(\vec{r}_{1},\vec{r}_{0};z)}{L}
=
\frac{-2Z_0c}{\gamma ab}
\sum _{m,n\geq 0} 
   \frac{k_y}{k_c}
   \phi' _{mny}\left(\vec{r}_{1}\right)
   \phi' _{mnz}\left(\vec{r}_{0}\right)
e^{-\gamma k_c |z|}
,
\label{eq:PointChargeYWakeFunctionInWaveguide1}
\end{equation}
\label{eq:PointChargeWakeFunctionInWaveguide1}
\end{subequations}
where $\text{sgn}(z)$ denotes the sign function. Similar formulations of space-charge wake potentails were presented in Ref.~\cite{Nogales2012evaluation}.

Utilizing the fact of
\begin{equation}
\frac{1}{k^2+\beta^2\gamma^2k_c^2}
=
\int_0^\infty e^{-\left( k^2+\beta^2\gamma^2k_c^2 \right)t} dt
,
\label{eq:Integral1}
\end{equation}
Eqs.~(\ref{eq:PointChargeImpedanceInWaveguide1}) can be changed to the integral forms of
\begin{subequations}
\begin{equation}
\frac{Z_\parallel(\vec{r}_{1},\vec{r}_{0};k)}{L}
=
\frac{4iZ_0k}{ab}
\sum _{m,n\geq 0} 
\int_0^\infty dt e^{-\left( k^2+\beta^2\gamma^2k_c^2 \right)t}
   \phi' _{mnz}\left(\vec{r}_{1}\right)
   \phi' _{mnz}\left(\vec{r}_{0}\right)
,
\label{eq:PointChargeLongImpedanceInWaveguideInt1}
\end{equation}
\begin{equation}
\frac{Z_x(\vec{r}_{1},\vec{r}_{0};k)}{L}
=
\frac{-4Z_0\beta \kappa}{ab}
\sum _{m,n\geq 0} 
\int_0^\infty dt e^{-\left( k^2+\beta^2\gamma^2k_c^2 \right)t}
   k_x
   \phi' _{mnx}\left(\vec{r}_{1}\right)
   \phi' _{mnz}\left(\vec{r}_{0}\right)
,
\label{eq:PointChargeXImpedanceInWaveguideInt1}
\end{equation}
\begin{equation}
\frac{Z_y(\vec{r}_{1},\vec{r}_{0};k)}{L}
=
\frac{-4Z_0\beta \kappa}{ab}
\sum _{m,n\geq 0} 
\int_0^\infty dt e^{-\left( k^2+\beta^2\gamma^2k_c^2 \right)t}
   k_y
   \phi' _{mny}\left(\vec{r}_{1}\right)
   \phi' _{mnz}\left(\vec{r}_{0}\right)
.
\label{eq:PointChargeYImpedanceInWaveguideInt1}
\end{equation}
\label{eq:PointChargeImpedanceInWaveguideInt1}
\end{subequations}

Equations~(\ref{eq:PointChargeImpedanceInWaveguide1}) and~(\ref{eq:PointChargeImpedanceInWaveguideInt1}) are the Green's function forms of the impedance driven by a point charge in rectangular chambers. It is seen that they depend on the positions of both the source and test particles. Note that this formulation is obtained in the Cartesian coordinate system instead of the Cylindrical one. One can see that these formulae are fully differentiable with the transverse coordinates $x_\text{i}$ and $y_\text{i}$ ($\text{i}$=0,1). With respect to the position of the test particle, a simple relation between the longitudinal and transverse impedances is easily verified:
\begin{subequations}
\begin{equation}
Z_x(\vec{r}_{1},\vec{r}_{0};k) = \frac{i\beta\kappa}{k} \frac{\partial Z_\parallel (\vec{r}_{1},\vec{r}_{0};k)}{\partial x_1},
\end{equation}
\begin{equation}
Z_y(\vec{r}_{1},\vec{r}_{0};k) = \frac{i\beta\kappa}{k} \frac{\partial Z_\parallel (\vec{r}_{1},\vec{r}_{0};k)}{\partial y_1}.
\end{equation}
\label{eq:LongToTransImpedanceRelation}
\end{subequations}
This relation is quite general and it applies to most of the results found in this paper, including the impedance of a bunch distribution in the form of Eq.~(\ref{eq:Impedance_Bunch_General1}).

When the transverse dimensions of the chamber $a$ and $b$ are large enough, using the Euler-Maclaurin formula the summation over $m$ and $n$ in Eqs.~(\ref{eq:PointChargeImpedanceInWaveguideInt1}) can be replaced by the integrations over the transverse wavenumbers:
\begin{equation}
\sum _{m,n\geq 0}
\rightarrow 
\frac{ab}{\pi^2}
\int_0^\infty dk_x\int_0^\infty dk_y
.
\label{eq:SummationToIntegration}
\end{equation}
In the process there is a trick that should be employed when the summations over $m$ and $n$ are performed. Take Eq.~(\ref{eq:PointChargeLongImpedanceInWaveguideInt1}) as an example, the summation over $m$ should be expanded first as follows
\begin{equation}
\begin{split}
& \sum _{m\geq 0} 
\sin \left(k_x (x_1+a/2)\right)
\sin \left(k_x (x_0+a/2)\right)  \\
& =
\sum _{m\geq 0}^{m=even}
\sin \left(k_x x_1\right)
\sin \left(k_x x_0\right)
+
\sum _{m\geq 0}^{m=odd}
\cos \left(k_x x_1\right)
\cos \left(k_x x_0\right)
\end{split}
.
\label{eq:SineFunctionExpansion1}
\end{equation}
With $m$ even or odd numbers, the summation over $m$ should be replaced by
\begin{equation}
\sum _{m\geq 0}^{m=\text{even}\:\text{or}\:\text{odd}}
\rightarrow 
\frac{a}{2\pi}
\int_0^\infty dk_x
.
\label{eq:SummationToIntegrationX}
\end{equation}
The integration over the sine and cosine parts of Eq.~(\ref{eq:SineFunctionExpansion1}) contains divergent terms but they are canceled with each other. Finally, from Eqs.~(\ref{eq:PointChargeImpedanceInWaveguideInt1}) we obtain the following results:
\begin{subequations}
\begin{equation}
\frac{Z_\parallel(\vec{r}_{1},\vec{r}_{0};k)}{L}
=
\frac{iZ_0k}{4\pi\beta^2\gamma^2}
\int_0^\infty dt
\frac{1}{t}
e^{-\frac{d^2}{t}}
e^{-\frac{k^2t}{4\beta^2\gamma^2}}
,
\label{eq:PointChargeLongImpedanceInFreeSpace1}
\end{equation}
\begin{equation}
\frac{Z_x(\vec{r}_{1},\vec{r}_{0};k)}{L}
=
\frac{Z_0\kappa(x_1-x_0)}{2\pi\beta\gamma^2}
\int_0^\infty dt
\frac{1}{t^2}
e^{-\frac{d^2}{t}}
e^{-\frac{k^2t}{4\beta^2\gamma^2}}
,
\label{eq:PointChargeXImpedanceInFreeSpace1}
\end{equation}
\begin{equation}
\frac{Z_y(\vec{r}_{1},\vec{r}_{0};k)}{L}
=
\frac{Z_0\kappa(y_1-y_0)}{2\pi\beta\gamma^2}
\int_0^\infty dt
\frac{1}{t^2}
e^{-\frac{d^2}{t}}
e^{-\frac{k^2t}{4\beta^2\gamma^2}}
,
\label{eq:PointChargeYImpedanceInFreeSpace1}
\end{equation}
\label{eq:PointChargeImpedanceInFreeSpace1}
\end{subequations}
with $d=\sqrt{(x_1-x_0)^2+(y_1-y_0)^2}$ the transverse distance between the source and test particles. The integrals over $t$ in the above equations can be expressed by Bessel functions of the second kind as follows
\begin{subequations}
\begin{equation}
\frac{Z_\parallel(\vec{r}_{1},\vec{r}_{0};k)}{L}
=
\frac{iZ_0k}{2\pi\beta^2\gamma^2}
K_0\left(
\xi_d
\right)
,
\label{eq:PointChargeLongImpedanceInFreeSpace2}
\end{equation}
\begin{equation}
\frac{Z_x(\vec{r}_{1},\vec{r}_{0};k)}{L}
=
\frac{Z_0k\kappa(x_1-x_0)}{2\pi\beta^2\gamma^3d}
K_1\left(
\xi_d
\right)
,
\label{eq:PointChargeXImpedanceInFreeSpace2}
\end{equation}
\begin{equation}
\frac{Z_y(\vec{r}_{1},\vec{r}_{0};k)}{L}
=
\frac{Z_0k\kappa(y_1-y_0)}{2\pi\beta^2\gamma^3d}
K_1\left(
\xi_d
\right)
,
\label{eq:PointChargeYImpedanceInFreeSpace2}
\end{equation}
\label{eq:PointChargeImpedanceInFreeSpace2}
\end{subequations}
with $\xi_d=kd/(\beta\gamma)$. The above expressions in terms of modified Bessel functions are only valid for positive $k$. For negative frequencies, one should use the relations of $Z_\parallel(-k)=[Z_\parallel(k)]^*$ and $Z_{x,y}(-k)=-[Z_{x,y}(k)]^*$ in the condition that $\kappa$ is purely imaginary. The transverse impedance found here should be compared with the direct space-charge impedance as rigorously formulated in Refs.~\cite{ShobudaPRSTAB2007, Shobuda:2018bqc} (for example, see Eq.(27) in Ref.~\cite{Shobuda:2018bqc}). Finally, apply Eqs.~(\ref{eq:PointChargeImpedanceInFreeSpace1}) to Eqs.~(\ref{eq:WakeFunctionByImpedance1}), we can obtain the wake functions
\begin{subequations}
\begin{equation}
\frac{w_z(\vec{r},\vec{r}_{0};z)}{L}
=
\frac{Z_0c\gamma}{4\pi}
\frac{z}{\left[ d^2+\gamma^2z^2 \right]^{3/2}}
,
\label{eq:LongitudinalWakeFunctionPointChargeFreeSpace1}
\end{equation}
\begin{equation}
\frac{w_x(\vec{r},\vec{r}_{0};z)}{L}
=
\frac{Z_0c}{4\pi\gamma}
\frac{x_1-x_0}{\left[ d^2+\gamma^2z^2 \right]^{3/2}}
,
\label{eq:TransverseWakeFunctionXPointChargeFreeSpace1}
\end{equation}
\begin{equation}
\frac{w_y(\vec{r},\vec{r}_{0};z)}{L}
=
\frac{Z_0c}{4\pi\gamma}
\frac{y_1-y_0}{\left[ d^2+\gamma^2z^2 \right]^{3/2}}
,
\label{eq:TransverseWakeFunctionYPointChargeFreeSpace1}
\end{equation}
\label{eq:WakeFunctionPointChargeFreeSpace1}
\end{subequations}
with $z=v\tau$. The above expressions should be very familiar to the reader since they are linearly proportional to Green's function fields of a point charge with constant velocity. Equation~(\ref{eq:PointChargeLongImpedanceInFreeSpace2}) and Eq.~(\ref{eq:LongitudinalWakeFunctionPointChargeFreeSpace1}) were used in analysis of longitudinal space-charge microbunching starting from short noice in Ref.~\cite{HuangFEL2008}.

From Eqs.~(\ref{eq:PointChargeImpedanceInWaveguide1}), ~(\ref{eq:PointChargeImpedanceInFreeSpace2}) and ~(\ref{eq:WakeFunctionPointChargeFreeSpace1}), one can observe that the space-charge wake functions and  impedance of a point charge are singular when the test particle is approaching to the source particle.  Therefore, directly applying them to simulations of space charge effects might be challenging, though not impossible~\cite{HessPRSTAB2007}. In the following sections we will use the formulations of this section as a basis to derive the formulae of space-charge impedance and wake potentials with various bunch distributions.

\section{\label{sec:Longitudinal-wake-with-bunch-distribution}Longitudinal wake fields and impedance with bunch distributions}

Models of the longitudinal space-charge impedance with transverse beam distributions were examined by M. Venturini in Ref.~\cite{VenturiniPRSTAB2008}. This section is an extension of Venturini's work.

\subsection{\label{sec:Longitudinal_Impedance_Transverse_Gaussian_Beam}Beam with transverse Gaussian density}

Suppose an asymmetric beam with bi-Gaussian transverse distribution is given by
\begin{equation}
\rho_\perp(x,y)=
\frac{1}{2\pi\sigma_x\sigma_y}
e^{-\frac{(x-x_c)^2}{2\sigma_x^2}-\frac{(y-y_c)^2}{2\sigma_y^2}}
.
\label{eq:GaussianBeamDistribution}
\end{equation}
The beam center is located at $\vec{r}_{c}=(x_c,y_c)$. Applying Eqs.~(\ref{eq:GaussianBeamDistribution}) and~(\ref{eq:PointChargeLongImpedanceInWaveguide1}) to Eq.~(\ref{eq:Impedance_Bunch_General1}) yields the longitudinal impedance of
\begin{equation}
\frac{\mathcal{Z}_\parallel(\vec{r}_{1},k)}{L}
=
\frac{4iZ_0k}{ab}
\sum _{m,n\geq 0} 
\frac{
   \phi' _{mnz}\left(\vec{r}_{1}\right)
   \Phi' _{mnz}\left(\vec{r}_{c}\right)}
{k^2+\beta^2\gamma^2k_c^2}
,
\label{eq:GaussianBeamLongImpedanceInWaveguide1}
\end{equation}
where $\Phi' _{mnz}\left(\vec{r}_{c}\right)$ is defined by
\begin{equation}
\Phi' _{mnz}\left(\vec{r}_{c}\right)=
\int_{-\frac{a}{2}}^{\frac{a}{2}} dx' \int_{-\frac{b}{2}}^{\frac{b}{2}} dy'
\phi' _{mnz}\left(x', y'\right)
\rho(x',y')
.
\label{eq:GaussianBeamPhi1}
\end{equation}
Equation~(\ref{eq:GaussianBeamLongImpedanceInWaveguide1}) is already useful for simulation of space-charge effects with vacuum chambers. A basically similar treatment was given by J. Qiang in Refs.~\cite{QiangPRAB2017,QiangPRAB2018}. The integral of Eq.~(\ref{eq:GaussianBeamPhi1}) has analytic solution but is a little complicated and not convenient for the following calculations. Here we take the assumption that the transverse beam sizes are much smaller than the chamber size, e.g. $\sigma_x \ll a$ and $\sigma_y \ll b$. It means the beam is confined inside an area much smaller than the chamber's cross section. Then the integral limit can be extended from finite to infinite dimensions, resulting a simple solution of
\begin{equation}
\Phi' _{mnz}\left(\vec{r}_{c}\right)=
e^{-\frac{1}{2}\left( k_x^2\sigma_x^2 + k_y^2\sigma_y^2 \right)}
S_x(x_c) S_y(y_c)
.
\label{eq:GaussianBeamPhi2}
\end{equation}
Immediately the techniques discussed in the previous section can be used to calculate the impedance in free space for a beam with transverse bi-Gaussian distribution, resulting in
\begin{equation}
\frac{\mathcal{Z}_\parallel(\vec{r}_{1},k)}{L}
=
\frac{iZ_0k}{4\pi\beta^2\gamma^2}
\int_0^\infty dt
\frac{1}{(t+2\sigma_x^2)^{1/2}(t+2\sigma_y^2)^{1/2}}
e^{-\frac{(x_1-x_c)^2}{t+2\sigma_x^2}-\frac{(y_1-y_c)^2}{t+2\sigma_y^2}}
e^{-\frac{k^2t}{4\beta^2\gamma^2}}
.
\label{eq:BiGaussianBeamLongImpedanceInFreespace1}
\end{equation}
The above equation can also be derived from directly applying Eqs.~(\ref{eq:GaussianBeamDistribution}) and~(\ref{eq:PointChargeLongImpedanceInFreeSpace1}) to Eq.~(\ref{eq:Impedance_Bunch_General1}). For simplicity of discussions, Eq.~(\ref{eq:BiGaussianBeamLongImpedanceInFreespace1}) can be rewritten as
\begin{equation}
\frac{\mathcal{Z}_\parallel(\vec{r}_{1},k)}{L}
=
\frac{iZ_0}{4\pi\beta\gamma\sigma_x}
F_z(x_1,y_1,\alpha,k)
,
\label{eq:BiGaussianBeamLongImpedanceInFreeSpace2}
\end{equation}
with 
\begin{equation}
F_z(x,y,\alpha,k)
=
\xi_x\int_0^\infty dt
\frac{1}{(t+1)^{1/2}(t+\alpha^2)^{1/2}}
e^{-\frac{(x-x_c)^2}{2\sigma_x^2(t+1)}-\frac{(y-y_c)^2}{2\sigma_x^2(t+\alpha^2)}}
e^{-\frac{\xi^2_xt}{2}}
.
\label{eq:BiGaussianBeamLongImpedanceInFreeSpaceFz1}
\end{equation}
Here we define the transverse bunch aspect ratio $\alpha=\sigma_y/\sigma_x$ and the dimensionless frequency $\xi_x=k\sigma_x/(\beta\gamma)$. The convergence property of $F_z(x,y,\alpha,k)$ very depends on the wave number $k$, therefore in Eq.~(\ref{eq:BiGaussianBeamLongImpedanceInFreeSpaceFz1}) it is better to include $k$  so that $F_z$ converges when $k\to 0$. The exponential term of $e^{-\xi_x^2t/2}$ in Eq.~(\ref{eq:BiGaussianBeamLongImpedanceInFreeSpaceFz1}) plays a role of damping at high frequencies. By comparing the other two exponential terms including the transverse beam sizes, it defines a threshold frequency of
\begin{equation}
k_{th} = \frac{\sqrt{2}\beta\gamma}{\min[\sigma_x,\sigma_y]}
,
\label{eq:SpaceChargeFrequencyThresshold}
\end{equation}
where $\text{min}[\sigma_x,\sigma_y]$ indicates the minimum values of $\sigma_x$ and $\sigma_y$. When $k\gg k_{th}$, i.e. the case of high frequencies, the space charge impedance decreases exponentially.

The longitudinal impedance expressed by Eq.~(\ref{eq:BiGaussianBeamLongImpedanceInFreeSpaceFz1}) is not integrable in general. For the case of axis-symmetric beam of $\alpha=1$ (suppose $\sigma_x=\sigma_y=\sigma$), it can be expressed by
\begin{equation}
	F_z(x,y,1,k)=
	\xi_x e^{\xi_\sigma^2/2} \Gamma\left( 0,\xi_\sigma^2/2; A\xi_\sigma^2/2\right),
\end{equation}
with $A=\left[(x-x_c)^2+(y-y_c)^2 \right]/(2\sigma^2)$ and the generalized incomplete Gamma function~\cite{CHAUDHRY1996371} defined by
\begin{equation}
	\Gamma(\nu,x;b)=\int_x^\infty t^{\nu-1}e^{-t-b/t}dt.
\end{equation}
This function can be expanded in terms of other special functions~\cite{CHAUDHRY1996371}, but not very useful for fast evaluations of $F_z(x,y,1,k)$. For the special case of on-axis impedance, it reduces to the usual incomplete Gamma function as
\begin{equation}
F_z(0,0,1,k)
=
\xi_x e^{\xi_\sigma^2/2}\Gamma\left(0, \xi_\sigma^2/2 \right)
,
\label{eq:BiGaussianBeamLongImpedanceInWaveguideFz0}
\end{equation}
where $\xi_\sigma=k\sigma/(\beta\gamma)$, and the incomplete Gamma function $\Gamma(0,x)$ is defined as
\begin{equation}
\Gamma(0,x) =\Gamma(0,x;0)= \int_x^\infty \frac{e^{-t}}{t} dt
.
\end{equation}
Equation (\ref{eq:BiGaussianBeamLongImpedanceInWaveguideFz0}) is exactly the same as Eq. (15) in Ref.~\cite{VenturiniPRSTAB2008} where the exponential-integral functions was used instead. In the low-frequency limit $\xi_\sigma \ll 1$, from $\Gamma[0,x]\approx -\gamma_E-\ln x$ for small $x$ the asymptotic expression of Eq.~(\ref{eq:BiGaussianBeamLongImpedanceInWaveguideFz0}) can be found as
\begin{equation}
F_z(0,0,1,k)
\approx
-\xi_x \left(\ln{\frac{\xi_\sigma^2}{2}} + \gamma_E\right)
,
\label{eq:BiGaussianBeamLongImpedanceInWaveguideFz00}
\end{equation}
with $\gamma_E\approx 0.577216$ the Euler's constant. In the high-frequency limit $\xi_\sigma \gg 1$, from $\Gamma[0,x]\approx e^{-x}/x$ for large $x$, Eq.~(\ref{eq:BiGaussianBeamLongImpedanceInWaveguideFz0}) can be approximated by
\begin{equation}
	F_z(0,0,1,k)
	\approx
	2\xi_x/\xi_\sigma^2
	,
	\label{eq:BiGaussianBeamLongImpedanceHighFreqLimitFz0}
\end{equation}
giving the prediction of decay as $1/k$ as well known. For the case of asymmetric beam, an approximation with $\xi_\sigma=(\sigma_x+\sigma_y)/2$ gives an alternative model of the on-axis impedance as
\begin{equation}
F_z(0,0,\alpha,k)
=
\xi_x e^{\frac{\xi_x^2(1+\alpha)^2}{8}}\Gamma\left(0, \frac{\xi_x^2(1+\alpha)^2}{8} \right)
.
\label{eq:BiGaussianBeamLongImpedanceApproxInWaveguideFz0}
\end{equation}

\begin{figure}[htbp]
	\centering
	\includegraphics[width=10cm]{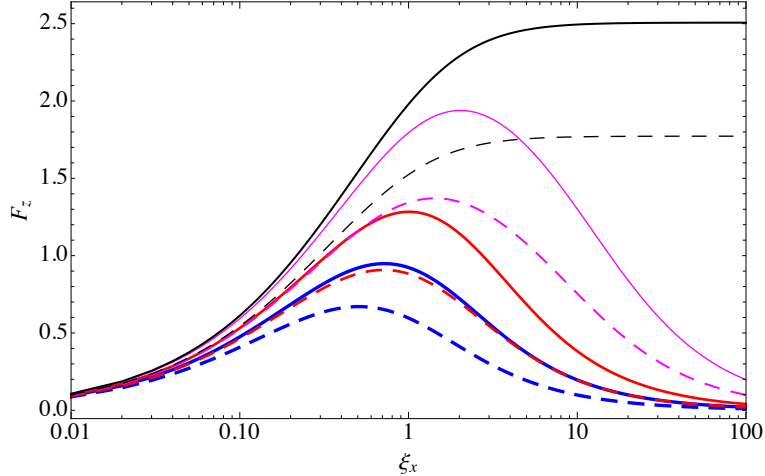}
	\caption{On-axis and average longitudinal impedance for bi-Gaussian beams. The solid and dashed lines are respectively from Eq.~(\ref{eq:BiGaussianBeamLongImpedanceInFreeSpaceFz1}) and Eq.~(\ref{eq:BiGaussianBeamAverageLongImpedanceInFreeSpaceFz1}). The blue, red, magenta, and black lines are for $\alpha=1$, 0.5, 0.1, and 0, respectively.}
	\label{fig:lsc_on-axis_comparison}
\end{figure}

An average of $F_z(x,y,\alpha,k)$ over the transverse distribution can be calculated from Eq.~(\ref{eq:Impedance_Average_Bunch_General1}), yielding
\begin{equation}
	\overline{F}_z(\alpha,k)
	=
	\xi_x\int_0^\infty dt
	\frac{1}{(t+2)^{1/2}(t+2\alpha^2)^{1/2}}
	e^{-\frac{\xi^2_xt}{2}}
	.
	\label{eq:BiGaussianBeamAverageLongImpedanceInFreeSpaceFz1}
\end{equation}
For a around beam with $\alpha=1$ and a flat beam with $\alpha=0$, the integral in the above equation can be expressed by incomplete Gamma function and modified Bessel function, respectively:
\begin{equation}
	\overline{F}_z(1,k)
	=
	\xi_x e^{\xi_\sigma^2}\Gamma\left(0, \xi_\sigma^2\right)
	,
	\label{eq:RoundGaussianBeamAverageLongImpedanceInFreeSpaceFz1}
\end{equation}
\begin{equation}
	\overline{F}_z(0,k)
	=
	\xi_x e^{\xi_x^2/2}
	K_0\left( \xi_x^2/2 \right)
	.
	\label{eq:FlatGaussianBeamAverageLongImpedanceInFreeSpaceFz1}
\end{equation}
The transverse beam aspect ratio is an important parameter in determining the scaling properties at high frequencies, see Fig.~\ref{fig:lsc_on-axis_comparison} for the on-axis and average longitudinal impedance as a function of dimensionless frequency $\xi_x$.

\begin{figure}[htbp]
	\centering
	\includegraphics[width=10cm]{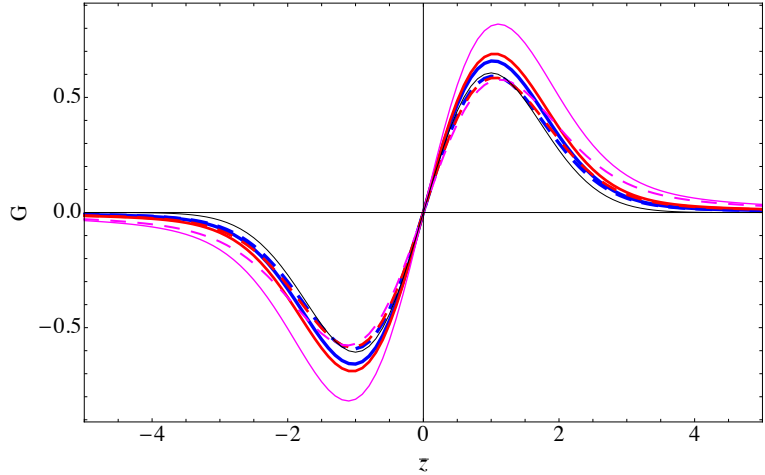}
	\caption{The distribution functions for the longitudinal space-charge wake potential of axis-symmetric Gaussian beams with $\alpha_{x,y} \ll 1$. The black solid line represents $G_0(\overline{z})$ defined by Eq.~(\ref{eq:FormFactorSimple1}). Other solid and dashed lines are respectively for Eq.~(\ref{eq:LongitudinalFormFactor1}) with $\vec{r}_1=0$ and Eq.~(\ref{eq:AverageLongitudinalFormFactor1}). The blue, red, and magenta lines are for $\alpha_x=\alpha_y=$0.001, 0.01, and 0.1, respectively.}
	\label{fig:lsc_wake_potential_comparison}
\end{figure}

\begin{figure}[htbp]
	\centering
	\includegraphics[width=10cm]{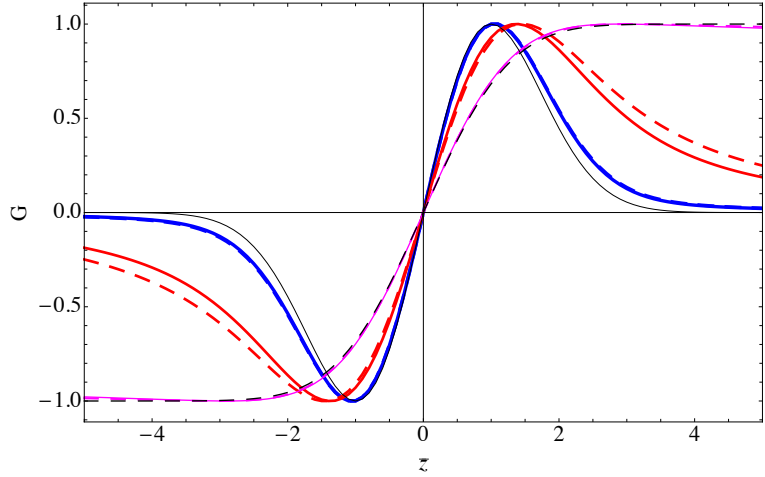}
	\caption{The distribution functions for the longitudinal space-charge wake potential of axis-symmetric Gaussian beams with small and large $\alpha_{x,y}$. The black solid and dashed lines are calculated from by Eqs.~(\ref{eq:FormFactorSimple1}) and~(\ref{eq:LongitudinalFormFactor2}). Other solid and dashed lines are respectively for Eq.~(\ref{eq:LongitudinalFormFactor1}) with $\vec{r}_1=0$ and Eq.~(\ref{eq:AverageLongitudinalFormFactor1}). The blue, red, and magenta lines are for $\alpha_x=\alpha_y=$0.01, 1, and 100, respectively. For convenience of comparison, the form factor $g$ for each line is numerically determined so that the maximum value is equal to 1.}
	\label{fig:lsc_wake_potential_comparison2}
\end{figure}

For a Gaussian bunch with length of $\sigma_z$, the wake potential (normalized by charge distribution) corresponding to Eq.~(\ref{eq:BiGaussianBeamLongImpedanceInFreespace1}) is
\begin{equation}
	\frac{W_z(\vec{r}_{1},z)}{L}
	=
	\frac{\gamma Z_0cz}{2\pi^{3/2}}
	\int_0^\infty dt
	\frac{1}{(t+2\sigma_x^2)^{1/2}(t+2\sigma_y^2)^{1/2}(t+2\gamma^2\sigma_z^2)^{3/2}}
	e^{-\frac{(x_1-x_c)^2}{t+2\sigma_x^2}-\frac{(y_1-y_c)^2}{t+2\sigma_y^2}-\frac{\gamma^2z^2}{t+2\gamma^2\sigma_z^2}}
	.
	\label{eq:BiGaussianBeamLongWakePotentialInFreespace1}
\end{equation}
Change the integral variable by $t=2\gamma^2\sigma_z^2t'$, we can rewrite the above equation as
\begin{equation}
	\frac{W_z(\vec{r}_{1},z)}{L}
	=
	\frac{Z_0cg}{4\pi \gamma^2}
	\frac{1}{\sqrt{2\pi}\sigma_z^2} G(\vec{r}_{1},\overline{z};\alpha_x,\alpha_y)
	,
	\label{eq:BiGaussianBeamLongWakePotentialInFreespace2}
\end{equation}
with the normalized longitudinal position $\overline{z}=z/\sigma_z$, the assumed form factor $g$, and the distribution function of
\begin{equation}
	G(\vec{r}_{1},\overline{z};\alpha_x,\alpha_y)=
	\frac{1}{g}
	\int_0^\infty dt
	\frac{\overline{z}}{(t+\alpha_x^2)^{1/2}(t+\alpha_y^2)^{1/2}(t+1)^{3/2}}
	e^{-\frac{(x_1-x_c)^2}{2\gamma^2\sigma_z^2(t+\alpha_x^2)}-\frac{(y_1-y_c)^2}{2\gamma^2\sigma_z^2(t+\alpha_y^2)}-\frac{\overline{z}^2}{2(t+1)}}
	,
	\label{eq:LongitudinalFormFactor1}
\end{equation}
where we define the transverse to longitudinal aspect ratios as $\alpha_x=\sigma_x/(\gamma\sigma_z)$ and $\alpha_y=\sigma_y/(\gamma\sigma_z)$ in the lab frame. The form factor $g=g(\alpha_x,\alpha_y)$ is a function of the beam aspect ratios to be determined. The above distribution function averaged over the transverse density is given by
\begin{equation}
	\overline{G}(\overline{z};\alpha_x,\alpha_y)=
	\frac{1}{\overline{g}}
	\int_0^\infty dt
	\frac{\overline{z}}{(t+2\alpha_x^2)^{1/2}(t+2\alpha_y^2)^{1/2}(t+1)^{3/2}}
	e^{-\frac{\overline{z}^2}{2(t+1)}}
	.
	\label{eq:AverageLongitudinalFormFactor1}
\end{equation}
A simple model of longitudinal space charge wake function is given by~\cite{Alex1993}
\begin{eqnarray}
	\frac{w_z(z)}{L}=
	\frac{Z_0cg_0}{4\pi \gamma^2} \delta'(z),
	\label{eq:LSCsimpleModel1}
\end{eqnarray}
where $g_0=1+\ln\frac{r_c}{r_b}$ is the form factor as a function of the transverse beam radius $r_b$ and chamber radius $r_c$. For a Gaussian bunch, it gives the wake potential of
\begin{eqnarray}
	\frac{W_z(z)}{L}=
	\frac{Z_0cg}{4\pi \gamma^2} \frac{1}{\sqrt{2\pi}\sigma_z^2}
	G_0(\overline{z})
	,
	\label{eq:LSCWakePotentialsimpleModel1}
\end{eqnarray}
with the simple distribution function of
\begin{equation}
	G_0(z)= \overline{z}e^{-\overline{z}^2/2}
	.
	\label{eq:FormFactorSimple1}
\end{equation}
We can use the low-frequency limit of Eq.~(\ref{eq:BiGaussianBeamLongImpedanceInWaveguideFz00}) with $\sigma=(\sigma_x+\sigma_y)/2$ for the longitudinal space-charge impedance to calculate the wake potential of a Gaussian bunch. Then by comparing the above simple model, we can determine the form factor as
\begin{equation}
	g(\alpha_x,\alpha_y)=4\ln2-2-2\ln(\alpha_x+\alpha_y)\ \text{for}\ \alpha_{x,y} \ll 1
	.
	\label{eq:Form_factor_low_frequency_limit}
\end{equation}
For the averaged impedance, there is
\begin{equation}
	\overline{g}=3\ln2-2-2\ln(\alpha_x+\alpha_y).
	\label{eq:FormFactorSimple2}
\end{equation}
We show the distribution functions $G_0(\overline{z})$, $\overline{G}(\overline{z};\alpha_x,\alpha_y)$, and $G(0,\overline{z};\alpha_x,\alpha_y)$ for three cases of axis-symmetric beams (i.e. $\alpha_x=\alpha_y=$0.001, 0.01, and 0.1) in Fig.~\ref{fig:lsc_wake_potential_comparison}, in which the form factors of Eqs.~(\ref{eq:Form_factor_low_frequency_limit}) and~(\ref{eq:FormFactorSimple2}) are used. 
In the limit of $\alpha_{x,y} \gg 1$, from Eq.~(\ref{eq:LongitudinalFormFactor1}) the on-axis impedance converges to
\begin{equation}
	G(0,\overline{z};\alpha_x,\alpha_y)=
	\text{efc} \left(\frac{\overline{z}}{\sqrt{2}}\right)
	,
	\label{eq:LongitudinalFormFactor2}
\end{equation}
with $g=\sqrt{2\pi}/(\alpha_x\alpha_y)$, and $\text{efc}(z)$ is the error function defined by $\text{efc}(x)=\frac{2}{\sqrt{\pi}}\int_0^ze^{-t^2}dt$. Equation~(\ref{eq:LongitudinalFormFactor2}) for the wake potential corresponds to the high-frequency impedance defined by Eq.~(\ref{eq:BiGaussianBeamLongImpedanceHighFreqLimitFz0}). The dependence of the distribution function $G$ on the beam aspect ratios is illustrated in Fig.~\ref{fig:lsc_wake_potential_comparison2} for both small and large $\alpha_{x,y}$.

\subsection{\label{sec:Longitudinal_Impedance_Transverse_delta_ring_Beam}$\delta$-ring beam}

Suppose a $\delta$-ring beam distribution defined in polar coordinates as
\begin{equation}
\rho_\perp(r,\theta)
=
\frac{1}{2\pi r_b}\delta(r-r_b)
,
\label{eq:delta-ring-distribution}
\end{equation}
with $r_b$ the ring's radius in the transverse plane. Apply Eqs.~(\ref{eq:delta-ring-distribution}) and~(\ref{eq:PointChargeLongImpedanceInFreeSpace2}) to Eq.~(\ref{eq:Impedance_Bunch_General1}) with the coordinates translated to polar coordinate system. Also consider the independence of the impedance to the angle variable of the test particle, we can obtain the radial dependent longitudinal impedance of
\begin{equation}
\mathcal{Z}_\parallel(r;k)
=
\frac{ikZ_0}{4\pi^2\beta^2\gamma^2}
\int_0^{2\pi}d\theta
K_0 \left( \sqrt{\xi^2_r+\xi_b^2-2\xi_r\xi_b\cos{\theta}} \right)
,
\label{eq:LongImpedance_delta_ring1}
\end{equation}
with $\xi_r=kr/(\beta\gamma)$ and $\xi_b=kr_b/(\beta\gamma)$ the dimensionless frequencies. The integration over $\theta$ in the above equation can be performed with the help of Graf's addition theorem for the modified Bessel functions~\cite{Watson1995treatise}, which indicates
\begin{equation}
K_\nu\left( \varpi \right) \cos{\nu\psi} 
=
\sum_{m=-\infty}^\infty K_{\nu+m}(Z)I_m(z) \cos{m\theta} 
,
\label{eq:Graf_addition_theorem}
\end{equation}
with $\varpi=\sqrt{Z^2+z^2-2Zz\cos{\theta}}$, $\cos{\psi}=(Z-z\cos{\theta})/\varpi$, and $\sin{\psi}=z\sin{\theta}/\varpi$. Equation~(\ref{eq:Graf_addition_theorem}) is valid for $|ze^{\pm i\theta}|<|Z|$. For $K_0(\varpi)$, only the $m=0$ term is necessarily counted in the integration over $\theta$ in Eq.~(\ref{eq:LongImpedance_delta_ring1}). Taking into account the validity condition of Eq.~(\ref{eq:Graf_addition_theorem}), the result can be expressed by
\begin{equation}
\mathcal{Z}_\parallel(r;k)
=
\frac{ikZ_0}{2\pi\beta^2\gamma^2}
\left[
\Theta(\xi_r-\xi_b)K_0(\xi_r)I_0(\xi_b) + \Theta(\xi_b-\xi_r)K_0(\xi_b)I_0(\xi_r)
\right]
,
\label{eq:LongImpedance_delta_ring2}
\end{equation}
where $\Theta(x)$ is the Heaviside step function defined as $\Theta(x)=1$ if $x>0$, and $\Theta(x)=0$ if $x<0$. Equation~(\ref{eq:LongImpedance_delta_ring2}) is consistent with Eq. (35) of Ref.~\cite{WuLSC2008} which can be simplified by applying the recursive relation of Bessel functions as pointed out in Ref.~\cite{HALAVANAU2016144}.

\subsection{\label{sec:Longitudinal_Impedance_Transverse_Uniform_Round_Beam}Round beam with transverse uniform density}

\begin{figure}[htbp]
	\centering
	\includegraphics[width=10cm]{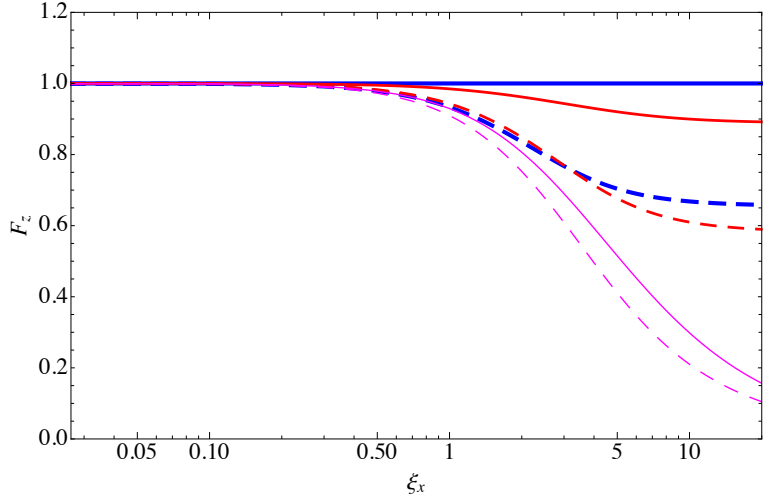}
	\caption{Comparison of the 1D models for on-axis longitudinal impedance. The solid and dashed lines are respectively from Eq.~(\ref{eq:BiGaussianBeamLongImpedanceApproxInWaveguideFz0}) and Eq.~(\ref{eq:LongImpedance_on-axis_Round-disk2}) normalized by $F_z(0,0,\alpha,k)$ as of Eq.~(\ref{eq:BiGaussianBeamLongImpedanceInFreeSpaceFz1}). The blue, red, and magenta lines are for $\alpha=1$, 0.5, and 0, respectively.}
	\label{fig:lsc_1d_comparison}
\end{figure}

Suppose a round-disk beam distribution defined in polar coordinates as
\begin{equation}
\rho_\perp(r,\theta)
=
\frac{1}{\pi r_b^2}\Theta(r_b-r)
,
\label{eq:Round-disk-distribution}
\end{equation}
with $r_b$ the radius of the disk. Apply Eqs.~(\ref{eq:Round-disk-distribution}) and~(\ref{eq:PointChargeLongImpedanceInFreeSpace2}) to Eq.~(\ref{eq:Impedance_Bunch_General1}) with translation to polar coordinates, or use Eq.~(\ref{eq:LongImpedance_delta_ring2}) as a Green's function as shown in Ref.~\cite{WuLSC2008}, we can obtain the radial dependent impedance of
\begin{equation}
\mathcal{Z}_\parallel(r;k)=
    \begin{cases}
      \frac{iZ_0}{\pi kr_b^2} \left[ 1-\xi_bK_1(\xi_b)I_0(\xi_r)\right], & \text{if}\ r<r_b, \\
      \frac{iZ_0}{\pi kr_b^2} \xi_bK_0(\xi_r)I_1(\xi_b), & \text{if}\ r>r_b.
    \end{cases}
\label{eq:LongImpedance_Round-disk1}
\end{equation}
The above equation is consistent with the result reported in Refs.~\cite{Rosenzweig1996-15,WangPRSTAB2015}. For a special case, the on-axis impedance is given by
\begin{equation}
	\mathcal{Z}_\parallel(0;k)=
	\frac{iZ_0}{4\pi \beta\gamma\sigma_x} F_z(0,k)
	,
	\label{eq:LongImpedance_on-axis_Round-disk1}
\end{equation}
with
\begin{equation}
	F_z(0;k)=
	\frac{4\xi_x}{\xi_b^2} \left[ 1-\xi_bK_1(\xi_b)\right]
	.
	\label{eq:LongImpedance_on-axis_Round-disk2}
\end{equation}
We factorize the impedance in the above form only for convenience of comparison as will be shown later. Equation~(\ref{eq:BiGaussianBeamLongImpedanceInWaveguideFz0}) with $\sigma=(\sigma_x+\sigma_y)/2$ and Eq.~(\ref{eq:LongImpedance_on-axis_Round-disk2}) with $r_b=1.747(\sigma_x+\sigma_y)/2$~\cite{VenturiniPRSTAB2008} are popular 1D models of longitudinal space-charge impedance for simulations of microbunching instability in free-electron lasers where usually it is valid to use bi-Gaussian distributions in the transverse directions (for an application of the second model, see Ref.~\cite{BorlandelegantCode}). We can compare these two 1D models of on-axis impedance with the exact expression of Eq.~(\ref{eq:BiGaussianBeamLongImpedanceInFreeSpaceFz1}). This is shown in Fig.~\ref{fig:lsc_1d_comparison}. One can see that at low frequencies (i.e. $\xi_x \ll 1$), all three models agree well with each other. For transversely asymmetric bi-Gaussian beams, at high frequencies (i.e. $\xi_x \gg 1$) the two approximate models underestimate the longitudinal space charge impedance. In general, the 1D model of Eq.~(\ref{eq:BiGaussianBeamLongImpedanceInWaveguideFz0}) is a better approximation than that of Eq.~(\ref{eq:LongImpedance_on-axis_Round-disk2}).

\begin{figure}[htbp]
	\centering
	\includegraphics[width=10cm]{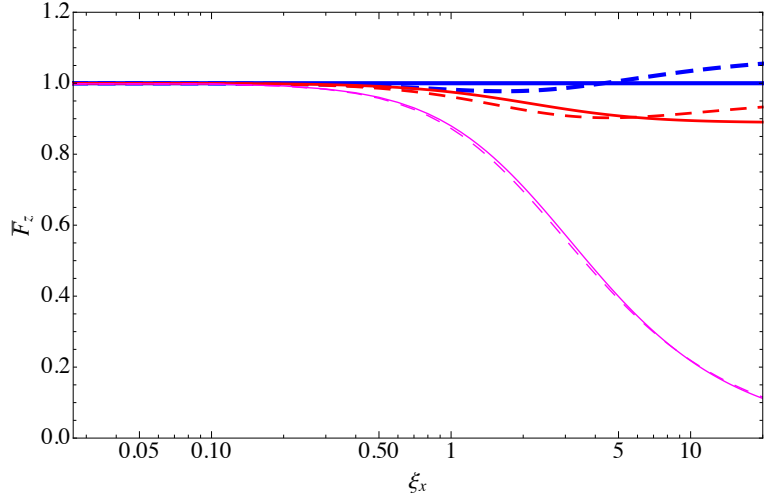}
	\caption{Comparison of the models for average longitudinal impedance. The solid and dashed lines are respectively from Eq.~(\ref{eq:RoundGaussianBeamAverageLongImpedanceInFreeSpaceFz1}) and Eq.~(\ref{eq:LongImpedance_average_Round-disk2}) normalized by $\overline{F}_z(\alpha,k)$ as of Eq.~(\ref{eq:BiGaussianBeamAverageLongImpedanceInFreeSpaceFz1}). Here prescriptions $r_b=1.924\sigma$ and $\sigma=(\sigma_x+\sigma_y)/2$ are used. The blue, red, and magenta lines are for $\alpha=1$, 0.5, and 0, respectively.}
	\label{fig:lsc_average_comparison}
\end{figure}

The average of Eq.~(\ref{eq:LongImpedance_Round-disk1}) over the beam distribution gives
\begin{equation}
\overline{\mathcal{Z}}_\parallel(k)=
\frac{iZ_0}{4\pi \beta\gamma\sigma_x} \overline{F}_z(k)
,
\label{eq:LongImpedance_average_Round-disk1}
\end{equation}
with
\begin{equation}
	\overline{F}_z(k)=
	\frac{4\xi_x}{\xi_b^2} \left[ 1-2K_1(\xi_b)I_1(\xi_b)\right]
	.
	\label{eq:LongImpedance_average_Round-disk2}
\end{equation}
Exactly this result was found in Ref.~\cite{VenturiniLSC2007}. The low-frequency approximation of Eq.~(\ref{eq:LongImpedance_average_Round-disk2}) can be found as
\begin{equation}
	\overline{F}_z(k) \approx
	2\xi_x \left( \frac{1}{4}-\gamma_E+\ln 2 - \ln \xi_b \right)
	.
	\label{eq:LongImpedance_average_Round-disk3}
\end{equation}
Similar to the treatment in Ref.~\cite{VenturiniPRSTAB2008}, we can equalize Eq.~(\ref{eq:LongImpedance_average_Round-disk3}) and the low-frequency asymptotic of Eq.~(\ref{eq:RoundGaussianBeamAverageLongImpedanceInFreeSpaceFz1}), then we obtain the equivalent radius of $r_b=2\sigma e^{(1-2\gamma_E)/4}\approx 1.924\sigma$. With this relation and $\sigma=(\sigma_x+\sigma_y)/2$, Eq.~(\ref{eq:LongImpedance_average_Round-disk2}) can be used to approximate the average longitudinal impedance for a bi-Gaussian beam. Note that the scaling factor 1.924 is slightly different from the value of 1.747 for the on-axis longitudinal impedance. Similar to the comparison of the models for on-axis impedance, we can also compare the models for average impedance as shown in Fig.~\ref{fig:lsc_average_comparison}. One can see that the approximate models have similar performance for different transverse beam aspect ratios. It is noteworthy that for flat beams with $\alpha \rightarrow 0$, all of the approximate models do not work well at high frequencies (i.e. $\xi_x \gg 1$).

\section{\label{sec:Transverse-wake-with-bunch-distribution}Transverse wake fields and impedance with bunch distributions}

With the longitudinal space charge impedance formulated the in the previous section, the transverse counterparts can be easily derived fro the relation of Eq.~(\ref{eq:LongToTransImpedanceRelation}).

\subsection{\label{sec:Transverse_Impedance_Transverse_Gaussian_Beam}Beam with transverse Gaussian density}

Applying Eqs.~(\ref{eq:GaussianBeamDistribution}),~(\ref{eq:PointChargeXImpedanceInWaveguide1}) and~(\ref{eq:PointChargeYImpedanceInWaveguide1}) to Eq.~(\ref{eq:Impedance_Bunch_General1}), or applying Eq.~(\ref{eq:GaussianBeamLongImpedanceInWaveguide1}) to Eq.~(\ref{eq:LongToTransImpedanceRelation}), we can obtain the transverse impedance of
\begin{subequations}
\begin{equation}
\frac{\mathcal{Z}_x(\vec{r}_1,k)}{L}
=
-\frac{4Z_0\beta\kappa}{ab}
\sum _{m,n\geq 0} 
\frac{
   k_x
   \phi' _{mnx}\left(\vec{r}_{1}\right)
   \Phi' _{mnz}\left(\vec{r}_{c}\right)}
{k^2+\beta^2\gamma^2(k_x^2+k_y^2)}
,
\label{eq:GaussianBeamXImpedanceInWaveguide1}
\end{equation}
\begin{equation}
\frac{\mathcal{Z}_y(\vec{r}_1,k)}{L}
=
-\frac{4Z_0\beta\kappa}{ab}
\sum _{m,n\geq 0} 
\frac{
   k_y
   \phi' _{mny}\left(\vec{r}_{1}\right)
   \Phi' _{mnz}\left(\vec{r}_{c}\right)}
{k^2+\beta^2\gamma^2(k_x^2+k_y^2)}
.
\label{eq:GaussianBeamYImpedanceInWaveguide1}
\end{equation}
\label{eq:GaussianBeamImpedanceInWaveguide1}
\end{subequations}
For the case of free space and transverse bi-Gaussian beam distribution, we can find
\begin{subequations}
\begin{equation}
\frac{\mathcal{Z}_x(\vec{r}_1,k)}{L}
=
\frac{Z_0\kappa}{2\pi\beta\gamma^2}
\int_0^\infty dt
\frac{x_1-x_c}{(t+2\sigma_x^2)^{3/2}(t+2\sigma_y^2)^{1/2}}
e^{-\frac{(x_1-x_c)^2}{t+2\sigma_x^2}-\frac{(y_1-y_c)^2}{t+2\sigma_y^2}}
e^{-\frac{k^2t}{4\beta^2\gamma^2}}
,
\label{eq:BiGaussianBeamXImpedanceInWaveguide1}
\end{equation}
\begin{equation}
\frac{\mathcal{Z}_y(\vec{r}_1,k)}{L}
=
\frac{Z_0\kappa}{2\pi\beta\gamma^2}
\int_0^\infty dt
\frac{y_1-y_c}{(t+2\sigma_x^2)^{1/2}(t+2\sigma_y^2)^{3/2}}
e^{-\frac{(x_1-x_c)^2}{t+2\sigma_x^2}-\frac{(y_1-y_c)^2}{t+2\sigma_y^2}}
e^{-\frac{k^2t}{4\beta^2\gamma^2}}
.
\label{eq:BiGaussianBeamYImpedanceInWaveguide1}
\end{equation}
\label{eq:BiGaussianBeamImpedanceInWaveguide1}
\end{subequations}
The above equations can be rewritten as follows
\begin{subequations}
\begin{equation}
\frac{\mathcal{Z}_x(\vec{r}_{1},k)}{L}
=
\frac{Z_0\kappa(x_1-x_c)}{4\pi\beta\gamma^2\sigma_x^2}
F_x(x_1,y_1,\alpha,k)
,
\label{eq:BiGaussianBeamXImpedanceInWaveguide2}
\end{equation}
\begin{equation}
\frac{\mathcal{Z}_y(\vec{r}_{1},k)}{L}
=
\frac{Z_0\kappa(y_1-y_c)}{4\pi\beta\gamma^2\sigma_x^2}
F_y(x_1,y_1,\alpha,k)
,
\label{eq:BiGaussianBeamYImpedanceInWaveguide2}
\end{equation}
\label{eq:BiGaussianBeamImpedanceInWaveguide2}
\end{subequations}
with 
\begin{subequations}
\begin{equation}
F_x(x,y,\alpha,k)
=
\int_0^\infty dt
\frac{1}{(t+1)^{3/2}(t+\alpha^2)^{1/2}}
e^{-\frac{(x-x_c)^2}{2\sigma_x^2(t+1)}-\frac{(y-y_c)^2}{2\sigma_x^2(t+\alpha^2)}}
e^{-\frac{\xi_x^2t}{2}}
,
\label{eq:BiGaussianBeamXImpedanceInWaveguideFx1}
\end{equation}
\begin{equation}
F_y(x,y,\alpha,k)
=
\int_0^\infty dt
\frac{1}{(t+1)^{1/2}(t+\alpha^2)^{3/2}}
e^{-\frac{(x-x_c)^2}{2\sigma_x^2(t+1)}-\frac{(y-y_c)^2}{2\sigma_x^2(t+\alpha^2)}}
e^{-\frac{\xi_x^2t}{2}}
.
\label{eq:BiGaussianBeamYImpedanceInWaveguideFy1}
\end{equation}
\label{eq:BiGaussianBeamImpedanceInWaveguideF1}
\end{subequations}
Similar to $F_z(x,y,\alpha,k)$, these functions are dimensionless. 

\begin{figure}[htbp]
	\centering
	\includegraphics[width=10cm]{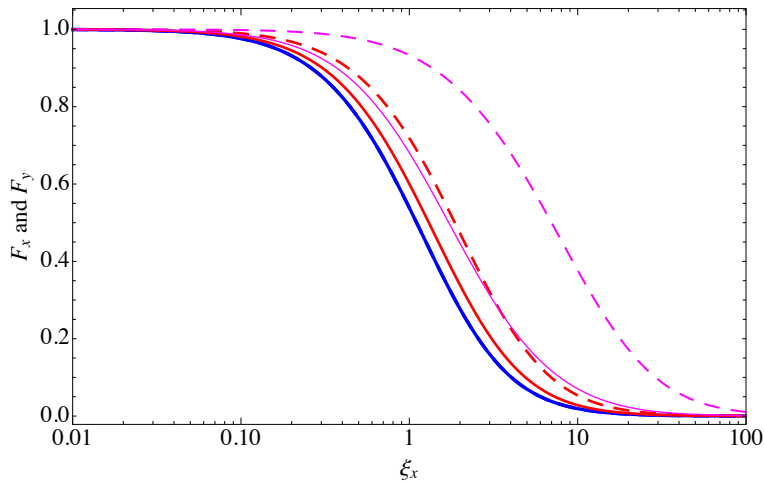}
	\caption{On-axis transverse impedance for bi-Gaussian beams. The solid and dashed lines are calculated from Eqs.~(\ref{eq:BiGaussianBeamImpedanceInWaveguideF1}) (normalized by Eqs.~(\ref{eq:BiGaussianBeamImpedanceInFreespaceF0})) for horizontal and vertical directions, respectively.  The blue, red, and magenta lines are for $\alpha=1$, 0.5, and 0.1, respectively. For $\alpha=1$ (blue), the two lines overlay each other.}
	\label{fig:tsc_on-axis_comparison}
\end{figure}

For a Gaussian bunch with length $\sigma_z$ in the longitudinal direction, the typical frequency of the beam spectrum is $k\sim 1/ \sigma_z$. If $1/\sigma_z \ll k_{th}$ is satisfied, the damping term of $e^{-\xi_x^2t/2}$ in Eqs.~(\ref{eq:BiGaussianBeamImpedanceInWaveguideF1}) can be neglected and the transverse space charge impedance can be approximated by
\begin{subequations}
\begin{equation}
F_x(x,y,\alpha,0)
=
\int_0^\infty dt
\frac{1}{(t+1)^{3/2}(t+\alpha^2)^{1/2}}
e^{-\frac{(x-x_c)^2}{2\sigma_x^2(t+1)}-\frac{(y-y_c)^2}{2\sigma_x^2(t+\alpha^2)}}
,
\label{eq:BiGaussianBeamXImpedanceInWaveguideFx2}
\end{equation}
\begin{equation}
F_y(x,y,\alpha,0)
=
\int_0^\infty dt
\frac{1}{(t+1)^{1/2}(t+\alpha^2)^{3/2}}
e^{-\frac{(x-x_c)^2}{2\sigma_x^2(t+1)}-\frac{(y-y_c)^2}{2\sigma_x^2(t+\alpha^2)}}
.
\label{eq:BiGaussianBeamYImpedanceInWaveguideFy2}
\end{equation}
\label{eq:BiGaussianBeamImpedanceInWaveguideF2}
\end{subequations}
The above equations are independent to frequency, so they represent the transverse space charge impedance for coasting beams~\cite{RuggieroCAS2005}. Suppose $\alpha<1$ and define new coordinates
\begin{equation}
	X=\frac{x_1-x_c}{\sqrt{2(\sigma_x^2-\sigma_y^2)}},\quad Y=\frac{y_1-y_c}{\sqrt{2(\sigma_x^2-\sigma_y^2)}},
	\label{eq:CoordinateTransform1}
\end{equation}
after proper substitutions of integral variables~\cite{XiaoTSC2007}, the low-frequency limit of transverse impedance can be formulated as
\begin{equation}
	\frac{\mathcal{Z}_x(\vec{r}_1,0)}{L}-i\frac{\mathcal{Z}_y(\vec{r}_1,0)}{L}
	=
	\frac{-iZ_0\kappa}{2\beta\gamma^2\sqrt{2\pi(\sigma_x^2-\sigma_y^2)}}\mathcal{W}(X,Y)
	,
	\label{eq:ErrorFunctionExpressionZxy1}
\end{equation}
with
\begin{equation}
	\mathcal{W}(X,Y)=
	w\left(X+iY\right) - e^{-B} w\left(\alpha X+iY/\alpha\right)
	.
	\label{eq:eq:ErrorFunctionExpressionW1}
\end{equation}
Here $B=X^2(1-\alpha^2)+Y^2(1/\alpha^2-1)$ and $w(x)$ represents the complex error function of
\begin{equation}
	w(x)=e^{-x^2} \left[1+\frac{2i}{\sqrt{\pi}}\int_0^xe^{t^2}dt \right].
	\label{eq:ComplexErrorFunction}	
\end{equation}
Equation~(\ref{eq:ErrorFunctionExpressionZxy1}) is exactly the Bassetti-Erskine formulation for beam-beam deflections of bi-Gaussian charge~\cite{Bassetti:1980by} and later proposed by A. Xiao \mbox{\emph{et al.\ }} for modeling of direct transverse space-charge forces. Suppose the transverse beam sizes are independent the longitudinal positions along a bunch, the transverse space-charge wake potential corresponding to Eq.~(\ref{eq:ErrorFunctionExpressionZxy1}) can be simply written as
\begin{equation}
	\frac{W_x(\vec{r}_1,z)}{L}-i\frac{W_y(\vec{r}_1,z)}{L}
	=
	\frac{-iZ_0\rho_z(z)}{2\beta\gamma^2\sqrt{2\pi(\sigma_x^2-\sigma_y^2)}}\mathcal{W}(X,Y)
	,
	\label{eq:ErrorFunctionExpressionWxy1}
\end{equation}
with $\rho_z(z)$ the longitudinal charge distribution normalized to unity. For a Gaussian bunch the above expression is valid only if $1/\sigma_z \ll k_{th}$. In the presence of microstructures with dimension of $\Delta z$ in the longitudinal profile, this criteria should be replaced by $1/\Delta z \ll k_{th}$ correspondingly for the applicability of Eqs.~(\ref{eq:BiGaussianBeamImpedanceInWaveguideF2}) and Eq.~(\ref{eq:ErrorFunctionExpressionWxy1}). For the case of on-axis impedance, we have
\begin{subequations}
\begin{equation}
F_x(0,0,\alpha,0)
=
\frac{2}{1+\alpha}
,
\label{eq:BiGaussianBeamXImpedanceInWaveguideFx0}
\end{equation}
\begin{equation}
F_y(0,0,\alpha,0)
=
\frac{2}{\alpha(1+\alpha)}
,
\label{eq:BiGaussianBeamYImpedanceInWaveguideFy0}
\end{equation}
\label{eq:BiGaussianBeamImpedanceInFreespaceF0}
\end{subequations}
representing the linear part, i.e. the popular one dimensional model, of the transverse space-charge impedance. For axis-symmetric beams, Eqs.~(\ref{eq:BiGaussianBeamImpedanceInWaveguideF2}) have analytical solution for the on-axis impedance as
\begin{equation}
F_\perp(0,0,1,k)
=
1-\frac{1}{2}\xi_\sigma^2e^{\frac{1}{2}\xi_\sigma^2}\Gamma \left(0,\frac{1}{2}\xi_\sigma^2 \right)
.
\label{eq:RoundGaussianBeamXImpedanceInFreespaceFx0}
\end{equation}
For illustration, we plot in Fig.~\ref{fig:tsc_on-axis_comparison} the functions of Eqs.~(\ref{eq:BiGaussianBeamImpedanceInWaveguideF1}) respectively normalized by Eqs.~(\ref{eq:BiGaussianBeamImpedanceInFreespaceF0}). It is obvious that the low-frequency limit of Eq.~(\ref{eq:ErrorFunctionExpressionZxy1}) for the transverse space-charge impedance is only valid for $\xi_x \ll 1$.
 
The transverse space charge impedance is nonlinear as a function of transverse coordinates. Usually the linear parts of Eqs.~(\ref{eq:BiGaussianBeamImpedanceInFreespaceF0}) are used for quick estimates of space charge effects. We can also calculate the average impedance over the transverse density. Using the horizontal impedance as an example, we first take derivative over $x_1$ and then do average over $x_1$ and $y_1$. Finally we obtain
\begin{subequations}
\begin{equation}
\overline{F}_x(\alpha, k)
=
\int_0^\infty dt
\frac{1}{(t+2)^{3/2}(t+2\alpha^2)^{1/2}}
e^{-\frac{\xi_x^2t}{2}}
,
\label{eq:BiGaussianBeamXAverageImpedanceInFreespaceFx1}
\end{equation}
\begin{equation}
\overline{F}_y(\alpha, k)
=
\int_0^\infty dt
\frac{1}{(t+2)^{1/2}(t+2\alpha^2)^{3/2}}
e^{-\frac{\xi_x^2t}{2}}
.
\label{eq:BiGaussianBeamYAverageImpedanceInFreespaceFy1}
\end{equation}
\label{eq:BiGaussianBeamAverageImpedanceInFreespaceF1}
\end{subequations}
Note that the above equations are not the direct average of Eqs.~(\ref{eq:BiGaussianBeamImpedanceInWaveguideF1}). Only for the round beam case there exists analytical solution of
\begin{equation}
\overline{F}_x(1, k)
=\overline{F}_y(1, k)
=
1-\frac{1}{2} \xi_\sigma^2 e^{\xi_\sigma^2} \Gamma\left(0, \xi_\sigma^2 \right)
.
\label{eq:RoundGaussianBeamXAverageImpedanceInFreespaceFxy1}
\end{equation}
The low-frequency limits of Eqs.~(\ref{eq:BiGaussianBeamAverageImpedanceInFreespaceF1}) give
\begin{subequations}
\begin{equation}
\overline{F}_x(\alpha, 0)
=
\frac{1}{1+\alpha}
,
\label{eq:BiGaussianBeamXAverageImpedanceDCInFreespaceFx1}
\end{equation}
\begin{equation}
\overline{F}_y(\alpha, 0)
=
\frac{1}{\alpha(1+\alpha)}
.
\label{eq:BiGaussianBeamYAverageImpedanceDCInFreespaceFy1}
\end{equation}
\label{eq:BiGaussianBeamAverageImpedanceDCInFreespaceF1}
\end{subequations}
One may notice that the average transverse space-charge impedance is half of the counterparts of the on-axis one for bi-Gaussian beams. This average effect is usually taken into account in the beam envelope equation with transverse space charge (for further illustrations, see Refs.~\cite{WanglerBook2008, ferrario2020injection}).

\subsection{\label{sec:Transverse_Impedance_Transverse_delta_ring_Beam}$\delta$-ring beam}

Apply Eqs.~(\ref{eq:delta-ring-distribution}),~(\ref{eq:PointChargeXImpedanceInFreeSpace2}) and~(\ref{eq:PointChargeYImpedanceInFreeSpace2}) to Eq.~(\ref{eq:Impedance_Bunch_General1}) with the coordinates translated to polar coordinate system, we can obtain the radial dependent transverse impedance of
\begin{equation}
	\mathcal{Z}_\perp(r;k)
	=
	\frac{kZ_0\kappa}{4\pi^2\beta^2\gamma^3}
	\int_0^{2\pi}d\theta
	\frac{\xi_r-\xi_b}{\sqrt{\xi^2_r+\xi_b^2-2\xi_r\xi_b\cos{\theta}}}
	K_1 \left( \sqrt{\xi^2_r+\xi_b^2-2\xi_r\xi_b\cos{\theta}} \right)
	.
	\label{eq:TransverseImpedance_delta_ring1}
\end{equation}
The integration over $\theta$ can be performed using the case of $\nu=1$ of Eq.~(\ref{eq:Graf_addition_theorem}). Again on the $m=0$ term survives and it results in
\begin{equation}
	\mathcal{Z}_\perp(r;k)
	=
	\frac{kZ_0\kappa}{2\pi\beta^2\gamma^3}
	\left[
	\Theta(\xi_r-\xi_b)K_1(\xi_r)I_0(\xi_b) - \Theta(\xi_b-\xi_r)K_0(\xi_b)I_1(\xi_r)
	\right]
	.
	\label{eq:TransverseImpedance_delta_ring2}
\end{equation}
The above equation can be easily obtained by applying the longitudinal impedance of Eq.~(\ref{eq:LongImpedance_delta_ring2}) to Eqs.~(\ref{eq:LongToTransImpedanceRelation}).

\subsection{\label{sec:Transverse_Impedance_Transverse_Uniform_Round_Beam}Round beam with transverse uniform density}

Apply Eqs.~(\ref{eq:Round-disk-distribution}) and~(\ref{eq:TransverseImpedance_delta_ring2}) to Eq.~(\ref{eq:Impedance_Bunch_General1}), or directly apply the longitudinal impedance in Eq.~(\ref{eq:LongImpedance_Round-disk1}) to Eqs.~(\ref{eq:LongToTransImpedanceRelation}), we can obtain the radial dependent transverse impedance of
\begin{equation}
	\mathcal{Z}_\perp(r;k)
	=
	\frac{Z_0\kappa}{\pi k\gamma r_b^2} \xi_b
	\left[
	\Theta(\xi_r-\xi_b)K_1(\xi_r)I_1(\xi_b) + \Theta(\xi_b-\xi_r)K_1(\xi_b)I_1(\xi_r)
	\right]
	.
	\label{eq:TransverseImpedance_Round_beam1}
\end{equation}
This is consistent with the results found in Ref.~\cite{Rosenzweig1996-15}. When $\xi_r \ll 1$ and $r<r_b$, it reduces to the linear dependent impedance over $r$:
\begin{equation}
	\mathcal{Z}_\perp(r;k)
	=
	\frac{Z_0\kappa k r}{2\pi \beta^2 \gamma^3} K_1(\xi_b)
	.
	\label{eq:TransverseImpedance_Round_beam2}
\end{equation}

\section{\label{sec:Summary} Summary}

By starting from the first principles defined in the Cartesian coordinate system, we have formulated explicitly the space-charge wake fields and impedance for a point charge in terms of eigenfunctions of rectangular chambers. Based on these results we derived various forms of wake fields and impedance in free space for the longitudinal and transverse space-charge effects. With only the rigid-beam and impulse approximation assumed, our formulations are valid for arbitrary beam velocity. Therefore, most of the equations in this paper are applicable to low energy beams.

On top of reproducing many of the analytical expressions already existing in the literature, our work also clearly show their applicable conditions. Furthermore, we found alternative models can be useful for modeling of space-charge effects. For examples, Eq.~(\ref{eq:BiGaussianBeamLongImpedanceInFreeSpaceFz1}) and Eqs.~(\ref{eq:BiGaussianBeamImpedanceInWaveguideF1}) can be used to model the space-charge forces of transverse bi-Gaussian beams (this is usually valid for 4th generation light sources). Their averages over the transverse density or approximated versions should be useful for fast simulations of space-charge effects. It worths noticing that when the average 1D model of Eq.~(\ref{eq:LongImpedance_average_Round-disk2}) (it is for round-disk beams) is applied to bi-Gaussian beams, the equivalent beam radius should be replaced by $r_b\approx 1.924(\sigma_x+\sigma_y)/2$ instead of the popular one (i.e. $r_b\approx 1.747(\sigma_x+\sigma_y)/2$) for on-axis model.

Besides the mode-expansion formulations of the space-charge wake fields and impedance in terms of elementary functions, the free-space results with various beam distributions can be used to calculate the corresponding results with parallel-plates or rectangular chambers using the image-charge method as reported in Ref.~\cite{WangPRSTAB2015}.

\begin{acknowledgments}
The author D.Z. would like to thank the cERL-FEL team at KEK, especially to N. Nakamura, T. Miyajima, O. Tanaka, and M. Shimada, for constant support and inspiring discussions. He is also thankful to Y. Shobuda in JAEA for useful discussions and suggestions. We also thank O. Tanaka for proofreading the paper.
\end{acknowledgments}

\nocite{*}

\bibliography{sc}

\begin{thebibliography}{31}%
\makeatletter
\providecommand \@ifxundefined [1]{%
 \@ifx{#1\undefined}
}%
\providecommand \@ifnum [1]{%
 \ifnum #1\expandafter \@firstoftwo
 \else \expandafter \@secondoftwo
 \fi
}%
\providecommand \@ifx [1]{%
 \ifx #1\expandafter \@firstoftwo
 \else \expandafter \@secondoftwo
 \fi
}%
\providecommand \natexlab [1]{#1}%
\providecommand \enquote  [1]{``#1''}%
\providecommand \bibnamefont  [1]{#1}%
\providecommand \bibfnamefont [1]{#1}%
\providecommand \citenamefont [1]{#1}%
\providecommand \href@noop [0]{\@secondoftwo}%
\providecommand \href [0]{\begingroup \@sanitize@url \@href}%
\providecommand \@href[1]{\@@startlink{#1}\@@href}%
\providecommand \@@href[1]{\endgroup#1\@@endlink}%
\providecommand \@sanitize@url [0]{\catcode `\\12\catcode `\$12\catcode
  `\&12\catcode `\#12\catcode `\^12\catcode `\_12\catcode `\%12\relax}%
\providecommand \@@startlink[1]{}%
\providecommand \@@endlink[0]{}%
\providecommand \url  [0]{\begingroup\@sanitize@url \@url }%
\providecommand \@url [1]{\endgroup\@href {#1}{\urlprefix }}%
\providecommand \urlprefix  [0]{URL }%
\providecommand \Eprint [0]{\href }%
\providecommand \doibase [0]{https://doi.org/}%
\providecommand \selectlanguage [0]{\@gobble}%
\providecommand \bibinfo  [0]{\@secondoftwo}%
\providecommand \bibfield  [0]{\@secondoftwo}%
\providecommand \translation [1]{[#1]}%
\providecommand \BibitemOpen [0]{}%
\providecommand \bibitemStop [0]{}%
\providecommand \bibitemNoStop [0]{.\EOS\space}%
\providecommand \EOS [0]{\spacefactor3000\relax}%
\providecommand \BibitemShut  [1]{\csname bibitem#1\endcsname}%
\let\auto@bib@innerbib\@empty
\bibitem [{\citenamefont {Qiang}\ \emph {et~al.}(2000)\citenamefont {Qiang},
  \citenamefont {Ryne}, \citenamefont {Habib},\ and\ \citenamefont
  {Decyk}}]{Qiang2000434}%
  \BibitemOpen
  \bibfield  {author} {\bibinfo {author} {\bibfnamefont {J.}~\bibnamefont
  {Qiang}}, \bibinfo {author} {\bibfnamefont {R.~D.}\ \bibnamefont {Ryne}},
  \bibinfo {author} {\bibfnamefont {S.}~\bibnamefont {Habib}},\ and\ \bibinfo
  {author} {\bibfnamefont {V.}~\bibnamefont {Decyk}},\ }\bibfield  {title}
  {\bibinfo {title} {An object-oriented parallel particle-in-cell code for beam
  dynamics simulation in linear accelerators},\ }\href
  {https://doi.org/https://doi.org/10.1006/jcph.2000.6570} {\bibfield
  {journal} {\bibinfo  {journal} {Journal of Computational Physics}\ }\textbf
  {\bibinfo {volume} {163}},\ \bibinfo {pages} {434 } (\bibinfo {year}
  {2000})}\BibitemShut {NoStop}%
\bibitem [{GPT()}]{GPTwebsite}%
  \BibitemOpen
  \href@noop {} {}\bibinfo {howpublished}
  {\url{http://www.pulsar.nl/gpt/}}\BibitemShut {NoStop}%
\bibitem [{\citenamefont {Borland}(2000)}]{BorlandelegantCode}%
  \BibitemOpen
  \bibfield  {author} {\bibinfo {author} {\bibfnamefont {M.}~\bibnamefont
  {Borland}},\ }\bibfield  {title} {\bibinfo {title} {{elegant: A Flexible
  SDDS-Compliant Code for Accelerator Simulation}},\ }in\ \href
  {https://doi.org/10.2172/761286} {\emph {\bibinfo {booktitle} {{6th
  International Computational Accelerator Physics Conference (ICAP 2000)}}}}\
  (\bibinfo {year} {2000})\BibitemShut {NoStop}%
\bibitem [{\citenamefont {Chao}(1993)}]{Alex1993}%
  \BibitemOpen
  \bibfield  {author} {\bibinfo {author} {\bibfnamefont {A.}~\bibnamefont
  {Chao}},\ }\bibinfo {title} {Physics of collective beam instabilities in high
  energy accelerators}\ (\bibinfo  {publisher} {Wiley},\ \bibinfo {year}
  {1993})\BibitemShut {NoStop}%
\bibitem [{\citenamefont {Ng}(2006)}]{NgBook2006}%
  \BibitemOpen
  \bibfield  {author} {\bibinfo {author} {\bibfnamefont {K.~Y.}\ \bibnamefont
  {Ng}},\ }\href {https://doi.org/10.1142/5835} {\emph {\bibinfo {title}
  {{Physics of intensity dependent beam instabilities}}}}\ (\bibinfo
  {publisher} {World Scientific},\ \bibinfo {address} {Hoboken, NJ},\ \bibinfo
  {year} {2006})\BibitemShut {NoStop}%
\bibitem [{\citenamefont {Shobuda}\ and\ \citenamefont
  {Chin}(2018)}]{Shobuda:2018bqc}%
  \BibitemOpen
  \bibfield  {author} {\bibinfo {author} {\bibfnamefont {Y.}~\bibnamefont
  {Shobuda}}\ and\ \bibinfo {author} {\bibfnamefont {Y.~H.}\ \bibnamefont
  {Chin}},\ }\bibfield  {title} {\bibinfo {title} {{Rigorous formulation of
  space-charge wake function and impedance by solving the three-dimensional
  Poisson equation}},\ }\href {https://doi.org/10.1038/s41598-018-30960-2}
  {\bibfield  {journal} {\bibinfo  {journal} {Sci. Rep.}\ }\textbf {\bibinfo
  {volume} {8}},\ \bibinfo {pages} {12805} (\bibinfo {year}
  {2018})}\BibitemShut {NoStop}%
\bibitem [{\citenamefont {Ng}(2004)}]{Ng:2004ky}%
  \BibitemOpen
  \bibfield  {author} {\bibinfo {author} {\bibfnamefont {K.}~\bibnamefont
  {Ng}},\ }\bibfield  {title} {\bibinfo {title} {{Space-charge impedances of
  beams with non-uniform transverse distributions}},\ }\href@noop {} {\
  (\bibinfo {year} {2004})}\BibitemShut {NoStop}%
\bibitem [{\citenamefont {Wang}\ and\ \citenamefont
  {Li}(2015)}]{WangPRSTAB2015}%
  \BibitemOpen
  \bibfield  {author} {\bibinfo {author} {\bibfnamefont {L.}~\bibnamefont
  {Wang}}\ and\ \bibinfo {author} {\bibfnamefont {Y.}~\bibnamefont {Li}},\
  }\bibfield  {title} {\bibinfo {title} {Analysis of the longitudinal space
  charge impedance of a round uniform beam inside parallel plates and
  rectangular chambers},\ }\href
  {https://doi.org/10.1103/PhysRevSTAB.18.024201} {\bibfield  {journal}
  {\bibinfo  {journal} {Phys. Rev. ST Accel. Beams}\ }\textbf {\bibinfo
  {volume} {18}},\ \bibinfo {pages} {024201} (\bibinfo {year}
  {2015})}\BibitemShut {NoStop}%
\bibitem [{\citenamefont {Persichelli}\ \emph {et~al.}(2017)\citenamefont
  {Persichelli}, \citenamefont {Biancacci}, \citenamefont {Migliorati},
  \citenamefont {Palumbo},\ and\ \citenamefont
  {Vaccaro}}]{PersichelliPRAB2017}%
  \BibitemOpen
  \bibfield  {author} {\bibinfo {author} {\bibfnamefont {S.}~\bibnamefont
  {Persichelli}}, \bibinfo {author} {\bibfnamefont {N.}~\bibnamefont
  {Biancacci}}, \bibinfo {author} {\bibfnamefont {M.}~\bibnamefont
  {Migliorati}}, \bibinfo {author} {\bibfnamefont {L.}~\bibnamefont
  {Palumbo}},\ and\ \bibinfo {author} {\bibfnamefont {V.~G.}\ \bibnamefont
  {Vaccaro}},\ }\bibfield  {title} {\bibinfo {title} {Electromagnetic fields
  and green's functions in elliptical vacuum chambers},\ }\href
  {https://doi.org/10.1103/PhysRevAccelBeams.20.101004} {\bibfield  {journal}
  {\bibinfo  {journal} {Phys. Rev. Accel. Beams}\ }\textbf {\bibinfo {volume}
  {20}},\ \bibinfo {pages} {101004} (\bibinfo {year} {2017})}\BibitemShut
  {NoStop}%
\bibitem [{\citenamefont {Migliorati}\ \emph {et~al.}(2018)\citenamefont
  {Migliorati}, \citenamefont {Biancacci}, \citenamefont {Masullo},
  \citenamefont {Palumbo},\ and\ \citenamefont {Vaccaro}}]{MiglioratiPRAB2018}%
  \BibitemOpen
  \bibfield  {author} {\bibinfo {author} {\bibfnamefont {M.}~\bibnamefont
  {Migliorati}}, \bibinfo {author} {\bibfnamefont {N.}~\bibnamefont
  {Biancacci}}, \bibinfo {author} {\bibfnamefont {M.~R.}\ \bibnamefont
  {Masullo}}, \bibinfo {author} {\bibfnamefont {L.}~\bibnamefont {Palumbo}},\
  and\ \bibinfo {author} {\bibfnamefont {V.~G.}\ \bibnamefont {Vaccaro}},\
  }\bibfield  {title} {\bibinfo {title} {Space charge impedance and
  electromagnetic fields in elliptical vacuum chambers},\ }\href
  {https://doi.org/10.1103/PhysRevAccelBeams.21.124201} {\bibfield  {journal}
  {\bibinfo  {journal} {Phys. Rev. Accel. Beams}\ }\textbf {\bibinfo {volume}
  {21}},\ \bibinfo {pages} {124201} (\bibinfo {year} {2018})}\BibitemShut
  {NoStop}%
\bibitem [{\citenamefont {Palumbo}\ \emph {et~al.}(1994)\citenamefont
  {Palumbo}, \citenamefont {Vaccaro},\ and\ \citenamefont
  {Zobov}}]{Palumbo1994}%
  \BibitemOpen
  \bibfield  {author} {\bibinfo {author} {\bibfnamefont {L.}~\bibnamefont
  {Palumbo}}, \bibinfo {author} {\bibfnamefont {V.~G.}\ \bibnamefont
  {Vaccaro}},\ and\ \bibinfo {author} {\bibfnamefont {M.}~\bibnamefont
  {Zobov}},\ }\href@noop {} {\emph {\bibinfo {title} {Wake Fields and
  Impedance}}},\ \bibinfo {type} {Tech. Rep.}\ \bibinfo {number}
  {LNF-94/041(P)}\ (\bibinfo {year} {September, 1994})\ \bibinfo {note} {``CAS
  - CERN Accelerator School: 5th Advanced Accelerator Physics
  Course''}\BibitemShut {NoStop}%
\bibitem [{\citenamefont {Chao}(2002)}]{ChaoNotes2002}%
  \BibitemOpen
  \bibfield  {author} {\bibinfo {author} {\bibfnamefont {A.}~\bibnamefont
  {Chao}},\ }\href@noop {} {\emph {\bibinfo {title} {Lecture Notes on Topics in
  Accelerator Physics}}}\ (\bibinfo {year} {November 2002})\ \bibinfo {note}
  {``SLAC-PUB-9574"}\BibitemShut {NoStop}%
\bibitem [{\citenamefont {Zhou}(2011)}]{ZhouThesis2011}%
  \BibitemOpen
  \bibfield  {author} {\bibinfo {author} {\bibfnamefont {D.}~\bibnamefont
  {Zhou}},\ }\emph {\bibinfo {title} {Coherent Synchrotron Radiation and
  Microwave Instability in Electron Storage Rings}},\ \href@noop {} {\bibinfo
  {type} {{Ph.D.} thesis}},\ \bibinfo  {school} {Graduate University for
  Advanced Studies, Japan} (\bibinfo {year} {2011})\BibitemShut {NoStop}%
\bibitem [{\citenamefont {Hess}\ \emph {et~al.}(2007)\citenamefont {Hess},
  \citenamefont {Park},\ and\ \citenamefont {Bolton}}]{HessPRSTAB2007}%
  \BibitemOpen
  \bibfield  {author} {\bibinfo {author} {\bibfnamefont {M.}~\bibnamefont
  {Hess}}, \bibinfo {author} {\bibfnamefont {C.~S.}\ \bibnamefont {Park}},\
  and\ \bibinfo {author} {\bibfnamefont {D.}~\bibnamefont {Bolton}},\
  }\bibfield  {title} {\bibinfo {title} {Green's function based space-charge
  field solver for electron source simulations},\ }\href
  {https://doi.org/10.1103/PhysRevSTAB.10.054201} {\bibfield  {journal}
  {\bibinfo  {journal} {Phys. Rev. ST Accel. Beams}\ }\textbf {\bibinfo
  {volume} {10}},\ \bibinfo {pages} {054201} (\bibinfo {year}
  {2007})}\BibitemShut {NoStop}%
\bibitem [{\citenamefont {Nogales}\ \emph {et~al.}(2012)\citenamefont
  {Nogales}, \citenamefont {Marini}, \citenamefont {Mart{\'\i}nez},
  \citenamefont {Melc{\'o}n}, \citenamefont {Pereira}, \citenamefont {Esbert},
  \citenamefont {Soto}, \citenamefont {Cogollos},\ and\ \citenamefont
  {Raboso}}]{Nogales2012evaluation}%
  \BibitemOpen
  \bibfield  {author} {\bibinfo {author} {\bibfnamefont {M.~J.}\ \bibnamefont
  {Nogales}}, \bibinfo {author} {\bibfnamefont {S.}~\bibnamefont {Marini}},
  \bibinfo {author} {\bibfnamefont {B.~G.}\ \bibnamefont {Mart{\'\i}nez}},
  \bibinfo {author} {\bibfnamefont {A.~{\'A}.}\ \bibnamefont {Melc{\'o}n}},
  \bibinfo {author} {\bibfnamefont {F.~Q.}\ \bibnamefont {Pereira}}, \bibinfo
  {author} {\bibfnamefont {V.~B.}\ \bibnamefont {Esbert}}, \bibinfo {author}
  {\bibfnamefont {P.}~\bibnamefont {Soto}}, \bibinfo {author} {\bibfnamefont
  {S.}~\bibnamefont {Cogollos}},\ and\ \bibinfo {author} {\bibfnamefont
  {D.}~\bibnamefont {Raboso}},\ }\bibfield  {title} {\bibinfo {title}
  {Evaluation of time domain electromagnetic fields radiated by constant
  velocity moving particles traveling along an arbitrarily shaped cross-section
  waveguide using frequency domain green's functions},\ }\href@noop {}
  {\bibfield  {journal} {\bibinfo  {journal} {Radio Science}\ }\textbf
  {\bibinfo {volume} {47}},\ \bibinfo {pages} {1} (\bibinfo {year}
  {2012})}\BibitemShut {NoStop}%
\bibitem [{\citenamefont {Shobuda}\ \emph {et~al.}(2007)\citenamefont
  {Shobuda}, \citenamefont {Chin},\ and\ \citenamefont
  {Takata}}]{ShobudaPRSTAB2007}%
  \BibitemOpen
  \bibfield  {author} {\bibinfo {author} {\bibfnamefont {Y.}~\bibnamefont
  {Shobuda}}, \bibinfo {author} {\bibfnamefont {Y.~H.}\ \bibnamefont {Chin}},\
  and\ \bibinfo {author} {\bibfnamefont {K.}~\bibnamefont {Takata}},\
  }\bibfield  {title} {\bibinfo {title} {Coupling impedances of a gap in vacuum
  chamber},\ }\href {https://doi.org/10.1103/PhysRevSTAB.10.044403} {\bibfield
  {journal} {\bibinfo  {journal} {Phys. Rev. ST Accel. Beams}\ }\textbf
  {\bibinfo {volume} {10}},\ \bibinfo {pages} {044403} (\bibinfo {year}
  {2007})}\BibitemShut {NoStop}%
\bibitem [{\citenamefont {Ratner}\ \emph {et~al.}(2008)\citenamefont {Ratner},
  \citenamefont {Chao},\ and\ \citenamefont {Huang}}]{HuangFEL2008}%
  \BibitemOpen
  \bibfield  {author} {\bibinfo {author} {\bibfnamefont {D.}~\bibnamefont
  {Ratner}}, \bibinfo {author} {\bibfnamefont {A.}~\bibnamefont {Chao}},\ and\
  \bibinfo {author} {\bibfnamefont {Z.}~\bibnamefont {Huang}},\ }\href@noop {}
  {\emph {\bibinfo {title} {Three-dimensional analysis of longitudinal space
  charge microbunching starting from shot noise}}},\ \bibinfo {type} {Tech.
  Rep.}\ (\bibinfo  {institution} {Stanford Linear Accelerator Center (SLAC)},\
  \bibinfo {year} {2008})\ \bibinfo {note} {``SLAC-PUB-13392"}\BibitemShut
  {NoStop}%
\bibitem [{\citenamefont {Venturini}(2008)}]{VenturiniPRSTAB2008}%
  \BibitemOpen
  \bibfield  {author} {\bibinfo {author} {\bibfnamefont {M.}~\bibnamefont
  {Venturini}},\ }\bibfield  {title} {\bibinfo {title} {Models of longitudinal
  space-charge impedance for microbunching instability},\ }\href
  {https://doi.org/10.1103/PhysRevSTAB.11.034401} {\bibfield  {journal}
  {\bibinfo  {journal} {Phys. Rev. ST Accel. Beams}\ }\textbf {\bibinfo
  {volume} {11}},\ \bibinfo {pages} {034401} (\bibinfo {year}
  {2008})}\BibitemShut {NoStop}%
\bibitem [{\citenamefont {Qiang}(2017)}]{QiangPRAB2017}%
  \BibitemOpen
  \bibfield  {author} {\bibinfo {author} {\bibfnamefont {J.}~\bibnamefont
  {Qiang}},\ }\bibfield  {title} {\bibinfo {title} {Symplectic multiparticle
  tracking model for self-consistent space-charge simulation},\ }\href
  {https://doi.org/10.1103/PhysRevAccelBeams.20.014203} {\bibfield  {journal}
  {\bibinfo  {journal} {Phys. Rev. Accel. Beams}\ }\textbf {\bibinfo {volume}
  {20}},\ \bibinfo {pages} {014203} (\bibinfo {year} {2017})}\BibitemShut
  {NoStop}%
\bibitem [{\citenamefont {Qiang}(2018)}]{QiangPRAB2018}%
  \BibitemOpen
  \bibfield  {author} {\bibinfo {author} {\bibfnamefont {J.}~\bibnamefont
  {Qiang}},\ }\bibfield  {title} {\bibinfo {title} {Symplectic particle-in-cell
  model for space-charge beam dynamics simulation},\ }\href
  {https://doi.org/10.1103/PhysRevAccelBeams.21.054201} {\bibfield  {journal}
  {\bibinfo  {journal} {Phys. Rev. Accel. Beams}\ }\textbf {\bibinfo {volume}
  {21}},\ \bibinfo {pages} {054201} (\bibinfo {year} {2018})}\BibitemShut
  {NoStop}%
\bibitem [{\citenamefont {Chaudhry}\ \emph {et~al.}(1996)\citenamefont
  {Chaudhry}, \citenamefont {Temme},\ and\ \citenamefont
  {Veling}}]{CHAUDHRY1996371}%
  \BibitemOpen
  \bibfield  {author} {\bibinfo {author} {\bibfnamefont {M.}~\bibnamefont
  {Chaudhry}}, \bibinfo {author} {\bibfnamefont {N.}~\bibnamefont {Temme}},\
  and\ \bibinfo {author} {\bibfnamefont {E.}~\bibnamefont {Veling}},\
  }\bibfield  {title} {\bibinfo {title} {Asymptotics and closed form of a
  generalized incomplete gamma function},\ }\href
  {https://doi.org/https://doi.org/10.1016/0377-0427(95)00018-6} {\bibfield
  {journal} {\bibinfo  {journal} {Journal of Computational and Applied
  Mathematics}\ }\textbf {\bibinfo {volume} {67}},\ \bibinfo {pages} {371 }
  (\bibinfo {year} {1996})}\BibitemShut {NoStop}%
\bibitem [{\citenamefont {Watson}(1995)}]{Watson1995treatise}%
  \BibitemOpen
  \bibfield  {author} {\bibinfo {author} {\bibfnamefont {G.}~\bibnamefont
  {Watson}},\ }\href@noop {} {\emph {\bibinfo {title} {A Treatise on the Theory
  of Bessel Functions}}},\ Cambridge Mathematical Library\ (\bibinfo
  {publisher} {Cambridge University Press},\ \bibinfo {year}
  {1995})\BibitemShut {NoStop}%
\bibitem [{\citenamefont {Wu}\ \emph {et~al.}(2008)\citenamefont {Wu},
  \citenamefont {Huang},\ and\ \citenamefont {Emma}}]{WuLSC2008}%
  \BibitemOpen
  \bibfield  {author} {\bibinfo {author} {\bibfnamefont {J.}~\bibnamefont
  {Wu}}, \bibinfo {author} {\bibfnamefont {Z.}~\bibnamefont {Huang}},\ and\
  \bibinfo {author} {\bibfnamefont {P.}~\bibnamefont {Emma}},\ }\bibfield
  {title} {\bibinfo {title} {Analytical analysis of longitudinal space charge
  effects for a bunched beam with radial dependence},\ }\href
  {https://doi.org/10.1103/PhysRevSTAB.11.040701} {\bibfield  {journal}
  {\bibinfo  {journal} {Phys. Rev. ST Accel. Beams}\ }\textbf {\bibinfo
  {volume} {11}},\ \bibinfo {pages} {040701} (\bibinfo {year}
  {2008})}\BibitemShut {NoStop}%
\bibitem [{\citenamefont {Halavanau}\ and\ \citenamefont
  {Piot}(2016)}]{HALAVANAU2016144}%
  \BibitemOpen
  \bibfield  {author} {\bibinfo {author} {\bibfnamefont {A.}~\bibnamefont
  {Halavanau}}\ and\ \bibinfo {author} {\bibfnamefont {P.}~\bibnamefont
  {Piot}},\ }\bibfield  {title} {\bibinfo {title} {Simulation of a cascaded
  longitudinal space charge amplifier for coherent radiation generation},\
  }\href {https://doi.org/https://doi.org/10.1016/j.nima.2016.03.002}
  {\bibfield  {journal} {\bibinfo  {journal} {Nuclear Instruments and Methods
  in Physics Research Section A: Accelerators, Spectrometers, Detectors and
  Associated Equipment}\ }\textbf {\bibinfo {volume} {819}},\ \bibinfo {pages}
  {144 } (\bibinfo {year} {2016})}\BibitemShut {NoStop}%
\bibitem [{\citenamefont {Rosenzweig}\ \emph {et~al.}(1996)\citenamefont
  {Rosenzweig}, \citenamefont {Pellegrini}, \citenamefont {Serafini},
  \citenamefont {Ternienden},\ and\ \citenamefont
  {Travish}}]{Rosenzweig1996-15}%
  \BibitemOpen
  \bibfield  {author} {\bibinfo {author} {\bibfnamefont {J.}~\bibnamefont
  {Rosenzweig}}, \bibinfo {author} {\bibfnamefont {C.}~\bibnamefont
  {Pellegrini}}, \bibinfo {author} {\bibfnamefont {L.}~\bibnamefont
  {Serafini}}, \bibinfo {author} {\bibfnamefont {C.}~\bibnamefont
  {Ternienden}},\ and\ \bibinfo {author} {\bibfnamefont {G.}~\bibnamefont
  {Travish}},\ }\href@noop {} {\emph {\bibinfo {title} {Space-Charge
  Oscillations in a Self-Modulated Electron Beam in Multi-Undulator
  Free-Electron Lasers}}},\ \bibinfo {type} {Tech. Rep.}\ (\bibinfo
  {institution} {DESY, Hamburg},\ \bibinfo {year} {1996})\ \bibinfo {note}
  {``TESLA FEL-Report 1996-15"}\BibitemShut {NoStop}%
\bibitem [{\citenamefont {Venturini}(2007)}]{VenturiniLSC2007}%
  \BibitemOpen
  \bibfield  {author} {\bibinfo {author} {\bibfnamefont {M.}~\bibnamefont
  {Venturini}},\ }\bibfield  {title} {\bibinfo {title} {{An effective
  longitudinal space-charge impedance model for beams with non-uniform and
  non-axissymmetric transverse density}}\ }\href
  {https://doi.org/10.2172/927331} {10.2172/927331} (\bibinfo {year}
  {2007})\BibitemShut {NoStop}%
\bibitem [{Rug(2005)}]{RuggieroCAS2005}%
  \BibitemOpen
  \href {https://doi.org/10.5170/CERN-2005-004} {\emph {\bibinfo {title} {{CAS
  - CERN Accelerator School : Basic Course on General Accelerator Physics:
  Loutraki, Greece 2 - 13 Oct 2000. CAS - CERN Accelerator School : Basic
  Course on General Accelerator Physics}}}},\ \bibinfo {organization} {CERN}\
  (\bibinfo  {publisher} {CERN},\ \bibinfo {address} {Geneva},\ \bibinfo {year}
  {2005})\ \bibinfo {note} {selected contributions}\BibitemShut {NoStop}%
\bibitem [{\citenamefont {Xiao}\ \emph {et~al.}(2007)\citenamefont {Xiao},
  \citenamefont {Borland}, \citenamefont {Emery}, \citenamefont {Wang},\ and\
  \citenamefont {Ng}}]{XiaoTSC2007}%
  \BibitemOpen
  \bibfield  {author} {\bibinfo {author} {\bibfnamefont {A.}~\bibnamefont
  {Xiao}}, \bibinfo {author} {\bibfnamefont {M.}~\bibnamefont {Borland}},
  \bibinfo {author} {\bibfnamefont {L.}~\bibnamefont {Emery}}, \bibinfo
  {author} {\bibfnamefont {Y.}~\bibnamefont {Wang}},\ and\ \bibinfo {author}
  {\bibfnamefont {K.}~\bibnamefont {Ng}},\ }\bibfield  {title} {\bibinfo
  {title} {{Direct Space Charge Calculation in Elegant and Its Application to
  the ILC Damping Ring}},\ }\href {https://doi.org/10.2172/921985} {\ ,\
  \bibinfo {pages} {3456} (\bibinfo {year} {2007})}\BibitemShut {NoStop}%
\bibitem [{\citenamefont {Bassetti}\ and\ \citenamefont
  {Erskine}(1980)}]{Bassetti:1980by}%
  \BibitemOpen
  \bibfield  {author} {\bibinfo {author} {\bibfnamefont {M.}~\bibnamefont
  {Bassetti}}\ and\ \bibinfo {author} {\bibfnamefont {G.}~\bibnamefont
  {Erskine}},\ }\bibfield  {title} {\bibinfo {title} {{Closed Expression for
  the Electrical Field of a Two-dimensional Gaussian Charge}},\ }\href@noop {}
  {\  (\bibinfo {year} {1980})}\BibitemShut {NoStop}%
\bibitem [{\citenamefont {Wangler}(2008)}]{WanglerBook2008}%
  \BibitemOpen
  \bibfield  {author} {\bibinfo {author} {\bibfnamefont {T.~P.}\ \bibnamefont
  {Wangler}},\ }\bibinfo {title} {Multiparticle dynamics with space charge},\
  in\ \href@noop {} {\emph {\bibinfo {booktitle} {RF Linear Accelerators}}}\
  (\bibinfo  {publisher} {John Wiley \& Sons, Ltd},\ \bibinfo {year} {2008})\
  Chap.~\bibinfo {chapter} {9}, pp.\ \bibinfo {pages} {282--340}\BibitemShut
  {NoStop}%
\bibitem [{\citenamefont {Ferrario}(2020)}]{ferrario2020injection}%
  \BibitemOpen
  \bibfield  {author} {\bibinfo {author} {\bibfnamefont {M.}~\bibnamefont
  {Ferrario}},\ }\href@noop {} {\bibinfo {title} {Injection, extraction and
  matching}} (\bibinfo {year} {2020}),\ \Eprint
  {https://arxiv.org/abs/2007.04102} {arXiv:2007.04102 [physics.acc-ph]}
  \BibitemShut {NoStop}%
\end{thebibliography}%

\end{document}